\documentclass[twocolumn, tighten, times, astrosymb]{aastex631}
\pdfoutput=1 

\usepackage{graphicx}
\usepackage[caption=false]{subfig}
\received{September 29, 2021}
\revised{December 22, 2021}
\accepted{December 28, 2021}

\submitjournal{\apj}

\shorttitle{New Clues to the Evolution of Dwarf Carbon Stars}
\shortauthors{Roulston et al.}

\turnoffediting

\def\Chandra{{\em Chandra}}
\def\Gaia{{\em Gaia }}

\newcommand{\aox}{\ifmmode{\alpha_{\mathrm{ox}}} \else $\alpha_{\mathrm{ox}}$\fi} 
\newcommand{\atoms}{\ifmmode{\mathrm{\,atoms~cm^{-2}}} \else \,atoms cm$^{-2}$\fi}
\newcommand{\ax}{\ifmmode{\alpha_x} \else $\alpha_x$\fi} 
\newcommand{\cmsq}{\ifmmode{\mathrm{cm^{-2}}} \else cm$^{-2}$\fi}
\newcommand{\degs}{\ifmmode ^{\circ}\else$^{\circ}$\fi}
\newcommand{\degsq}{\ifmmode {\mathrm{deg^2}} \else deg$^2$\fi}
\newcommand{\perdegsq}{\ifmmode {\mathrm{deg^{-2}}} \else deg$^{-2}$\fi}
\newcommand{\ew}{\ifmmode{W_{\lambda}} \else $W_{\lambda}$\fi}
\newcommand{\fbol}{\ifmmode f_{\mathrm{bol}} \else $f_{\mathrm{bol}}$\fi} 
\newcommand{\fcgs}{\ifmmode \mathrm{erg~cm^{-2}~s^{-1}}\else erg~cm$^{-2}$~s$^{-1}$\fi}
\newcommand{\flamcgs}{\ifmmode \mathrm{erg\,cm^{-2}\,s^{-1}\,\AA^{-1}}\else erg\,cm$^{-2}$\,s$^{-1}$\,\AA$^{-1}$)\fi}
\newcommand{\fnucgs}{\ifmmode {\mathrm{erg~cm^{-2}~s^{-1}~Hz^{-1}}}\else erg~cm$^{-2}$~s$^{-1}$~Hz$^{-1}$\fi}
\newcommand{\gax }{{\lower0.8ex\hbox{$\buildrel >\over\sim$}}}
\newcommand\Ha{\ifmmode {\mathrm H}\alpha \else H$\alpha$\fi}
\newcommand\Hb{\ifmmode {\mathrm H}\beta \else H$\beta$\fi}
\newcommand{\kms}{\ifmmode~{\mathrm{km~s}}^{-1}\else ~km~s$^{-1}~$\fi}
\newcommand{\lax }{{\lower0.8ex\hbox{$\buildrel <\over\sim$}}}
\newcommand{\lcgs}{\ifmmode \mathrm{erg~s^{-1}}\else erg~s$^{-1}$\fi}
\newcommand{\lnucgs}{\ifmmode erg~s^{-1}~Hz^{-1}\else erg~s$^{-1}$~Hz$^{-1}$\fi}

\newcommand{\logz}{\ifmmode{\mathrm{log}}~z \else log$~z$\fi}
\newcommand{\lo}{\ifmmode l_o \else $~l_o$\fi}
\newcommand{\Lo}{\ifmmode L_o \else $~L_o$\fi}
\newcommand{\lx}{\ifmmode l_x \else $~l_x$\fi}
\newcommand{\Lx}{\ifmmode L_x \else $~L_x$\fi}
\newcommand{\lbol}{\ifmmode L_{\mathrm{bol}} \else $L_{\mathrm{bol}}$\fi}
\newcommand{\Lbol}{\ifmmode L_{\mathrm{bol}} \else $L_{\mathrm{bol}}$\fi}
\newcommand{\LBol}{\ifmmode L_{\mathrm{bol}} \else $L_{\mathrm{bol}}$\fi}
\newcommand{\LEdd}{\ifmmode L_{\mathrm{Edd}} \else $L_{\mathrm{Edd}}$\fi}
\newcommand{\LxLbol}{\ifmmode L_x/L_{\mathrm{bol}} \else $L_x/L_{\mathrm{bol}}$\fi}
\newcommand{\rEdd}{\ifmmode L/L_{\mathrm{Edd}} \else $L/L_{\mathrm{Edd}}$\fi}
\newcommand{\REdd}{\ifmmode L/L_{\mathrm{Edd}} \else $L/L_{\mathrm{Edd}}$\fi}
\newcommand{\Rblr}{\ifmmode {R_{\mathrm BLR}} \else $R_{\mathrm BLR}$\fi}
\newcommand{\lamEdd}{\ifmmode \lambda_{\mathrm{Edd}} \else $\lambda_{\mathrm{Edd}}$\fi}
\newcommand{\mbh}{\ifmmode {M_{\rm BH}}\else${M_{\rm BH}}$\fi}
\newcommand{\Mbh}{\ifmmode {M_{\rm BH}}\else${M_{\rm BH}}$\fi}
\newcommand{\mdot}{\ifmmode \dot{m} \else $\dot{m}$\fi}
\newcommand{\mdote}{\ifmmode \dot{m}_{E} \else $\dot{m}_{E}$\fi}
\newcommand{\mone}{\ifmmode ^{-1}\else$^{-1}$\fi}
\newcommand{\msun}{\ifmmode {M_{\odot}}\else${M_{\odot}}$\fi}
\newcommand{\Msun}{\ifmmode {M_{\odot}}\else${M_{\odot}}$\fi}
\newcommand{\mtwo}{\ifmmode ^{-2}\else$^{-2}$\fi}
\newcommand{\Mvir}{\ifmmode {M_{\rm BH}^{\mathrm SE}}\else${M_{\rm BH}^{\mathrm SE}}$\fi}
\newcommand{\nhgal}{\ifmmode{ N_{H}^{Gal}} \else N$_{H}^{Gal}$\fi}
\newcommand{\nh}{\ifmmode{\mathrm N_{H}} \else N$_{H}$\fi}
\newcommand{\nhintr}{\ifmmode{ N_{H}^{intr}} \else N$_{H}^{intr}$\fi}
\newcommand{\nhtot}{\ifmmode{ N_{H}^{tot}} \else N$_{H}^{tot}$\fi}
\newcommand{\nhz}{\ifmmode{ N_{H}^z} \else N$_{H}^z$\fi}
\newcommand{\oi}{\ifmmode{\mathrm [O\,II]} \else [O\,II]\fi}
\newcommand{\oii}{\ifmmode{\mathrm [O\,II]} \else [O\,II]\fi}
\newcommand{\oiii}{\ifmmode{\mathrm [O\,III]} \else [O\,III]\fi}
\newcommand{\optebl}{\ifmmode L_{\rm 2500\,\AA} \else $~L_{\rm 2500\,\AA}$\fi}
\newcommand{\opteml}{\ifmmode l_{\mathrm{2500\,\AA}} \else $~l_{\mathrm{2500\,\AA}}$\fi}
\newcommand{\Teff}{\ifmmode T_{\mathrm{Eff}} \else $T_{\mathrm{Eff}}$\fi}
\newcommand{\xebl}{\ifmmode L_X \else $~L_X$\fi}
\newcommand{\xeml}{\ifmmode l_{\mathrm{2\,keV}} \else $~l_{\mathrm{2\,keV}}$\fi}

\def\geqsim{\lower.73ex\hbox{$\sim$}\llap{\raise.4ex\hbox{$>$}}$\,$}
\def\leqsim{\lower.73ex\hbox{$\sim$}\llap{\raise.4ex\hbox{$<$}}$\,$}

\newcommand{\umg}{\ifmmode{\mathrm{(}u-g\mathrm{)}} \else ($u-g$)\fi}
\newcommand{\gmr}{\ifmmode{\mathrm{(}g-r\mathrm{)}} \else ($g-r$)\fi}
\newcommand{\rmi}{\ifmmode{\mathrm{(}r-i\mathrm{)}} \else ($r-i$)\fi}
\newcommand{\gmi}{\ifmmode{\mathrm{(}g-i\mathrm{)}} \else ($g-i$)\fi}
\newcommand{\imz}{\ifmmode{\mathrm{(}i-z\mathrm{)}} \else ($i-z$)\fi}
\newcommand{\jmh}{\ifmmode{\mathrm{(}J-H\mathrm{)}} \else ($J-H$)\fi}
\newcommand{\hmk}{\ifmmode{\mathrm{(}H-K\mathrm{)}} \else ($H-K$)\fi}
\newcommand{\ctwo}{\ifmmode C_2 \else C$_2$\fi}

\begin{document}

\title{New Clues to the Evolution of Dwarf Carbon Stars From Their Variability and X-ray Emission}

\correspondingauthor{Benjamin Roulston}
\email{broulston@cfa.harvard.edu}

\author[0000-0002-9453-7735]{Benjamin R. Roulston}
\altaffiliation{SAO Predoctoral Fellow}
\affiliation{Center for Astrophysics $\vert$ Harvard \& Smithsonian, 60 Garden Street, Cambridge, MA 02138, USA}
\affiliation{Department of Astronomy, Boston University, 725 Commonwealth Avenue, Boston, MA 02215, USA}

\author[0000-0002-8179-9445]{Paul J. Green}
\affiliation{Center for Astrophysics $\vert$ Harvard \& Smithsonian, 60 Garden Street, Cambridge, MA 02138, USA}

\author[0000-0002-6752-2909]{Rodolfo Montez}
\affiliation{Center for Astrophysics $\vert$ Harvard \& Smithsonian, 60 Garden Street, Cambridge, MA 02138, USA}

\author[0000-0002-0201-8306]{Joseph Filippazzo}
\affiliation{Space Telescope Science Institute, 3700 San Martin Drive, Baltimore, MD 21218, USA}

\author[0000-0002-0210-2276]{Jeremy J. Drake}
\affiliation{Center for Astrophysics $\vert$ Harvard \& Smithsonian, 60 Garden Street, Cambridge, MA 02138, USA}

\author[0000-0002-2998-7940]{Silvia Toonen}
\affiliation{Anton Pannekoek Institute, University of Amsterdam, Science Park 904, 1098 XH Amsterdam, Netherlands}

\author[0000-0002-6404-9562]{Scott F. Anderson}
\affiliation{Department of Astronomy, University of Washington, Box 351580, Seattle, WA 98195, USA}

\author[0000-0002-3719-940X]{Michael Eracleous}
\affiliation{Department of Astronomy \& Astrophysics and Institute for Gravitation and the Cosmos, 525 Davey Laboratory, The Pennsylvania State University, University Park, PA 16802, USA}

\author[0000-0002-4948-7820]{Adam Frank}
\affiliation{Department of Physics and Astronomy, University of Rochester, Rochester, NY 14627-0171, USA}



\begin{abstract}
As main-sequence stars with C$>$O, dwarf carbon (dC) stars are never born alone but inherit carbon-enriched material from a former asymptotic giant branch (AGB) companion. In contrast to M dwarfs in post-mass transfer binaries, C$_2$ and/or CN molecular bands allow dCs to be identified with modest-resolution optical spectroscopy, even after the AGB remnant has cooled beyond detectability. Accretion of substantial material from the AGB stars should spin up the dCs, potentially causing a rejuvenation of activity detectable in X-rays.
Indeed, a few dozen dCs have recently been found to have photometric variability with periods under a day.  However, most of those are likely post-common-envelope binaries (PCEBs), spin-orbit locked by tidal forces, rather than solely spun-up by accretion.  Here, we study the X-ray properties of a sample of the five nearest known dCs with \Chandra.  Two are detected in X-rays, the only two for which we also detected short-period photometric variability.  We suggest that the coronal activity detected so far in dCs is attributable to rapid rotation due to tidal locking in short binary orbits after a common-envelope phase, late in the thermally pulsing (TP) phase of the former C-AGB primary (TP-AGB).
\end{abstract}



\section{Introduction}\label{sec:intro}

Dwarf carbon (dC) stars are \textit{main-sequence} stars that show molecular absorption bands of C, such as C$_2$, CN, and CH, in their optical spectra. Traditionally, carbon stars were thought to be enhanced intrinsically. Stars on the thermally pulsing (TP) phase of the asymptotic giant branch (AGB) experience shell He flashes. These He flashes cause strong convection in the intershell region, with resulting dredge-up \citep[the third dredge-up;][]{Iben1974} of He fusion products, namely carbon.
\added{As carbon-enriched material dredged up into the atmosphere of the AGB star then accretes onto its main sequence companion, that carbon preferentially binds with oxygen to form CO, and as C/O exceeds unity on the dwarf, the excess carbon is free to form the aforementioned molecules of C2, CN, and CH.}

This traditional explanation for C stars made it surprising when \citet{Dahn1977} found the first dwarf carbon star, G77-61. As dCs are main-sequence stars \deleted{(hydrogen core fusing)}, they could not have produced their own carbon, nor could they have experienced the third dredge-up necessary to bring this carbon to their envelopes. G77-61, and the hundreds of dC stars found since, must have been extrinsically enriched with carbon. \citet{Dahn1977} put forth a few theories for this extrinsic carbon enhancement, with the preferred method being binary mass transfer. 

\subsection{Binary Formation of dCs}

In the mass transfer scenario, the dC progenitor is in a binary system with a more massive star that evolved into a TP-AGB star. This TP-AGB star experienced intrinsic carbon enhancement as described above and became a giant C star itself. During the TP-AGB phase, stars can rapidly expand once C/O $> 1$, reaching radii of up to $800$\,R$_\sun$ \citep{Marigo2017} and can have slow, massive winds with mass loss rates of $\sim 10^{-7}$--$10^{-5}$\,M$_\sun$\,yr$^{-1}$ \citep{Hofner2018}.  This large, slow, carbon enhanced  wind can be accreted by the dC progenitor, bringing C/O $> 1$ and forming a dC. The TP-AGB then evolves further, expelling its envelope via a wind, leaving behind the CO core as a white dwarf (WD). The WD then cools over giga-year time scales, usually beyond detection in optical spectra.

Many studies have supported this binary mass transfer hypothesis. The first known dC, G77-61, was found to be a binary via radial velocity monitoring with a period of 245.5\,d \citep{Dearborn1986}.  Additionally, there have been almost a dozen ``smoking gun'' systems in which the WD is still visible in the optical spectrum, indicating more recent mass transer, i.e., a hot WD which has been cooling for a shorter time than in most dCs binaries \citep{Heber1993, Liebert1994, Green2013, Si2014}. \citet{Harris2018} found three dCs to be astrometric binaries with periods of 1.23\,yr, 3.21\,yr, and 11.35\,yr. Both \citet{Whitehouse2018} and \citet{Roulston2019} found radial velocity variations among a large sample of dCs, with the latter \replaced{finding}{inferring} a binary fraction near unity as would be expected from the mass transfer theory \citep[95\%;][]{Roulston2019}.
We note for context that the CH, Ba and the carbon-enhanced metal poor (CEMP-s) stars \citep{Lucatello2005}
 - mostly giants or subgiants - likely evolved from dC stars, and are better known than dCs only by virtue of their larger luminosities. 
 
It was traditionally thought that accretion of the mass needed for dCs to form would take place via Bondi-Hoyle-Lyttleton \citep{Hoyle1939, Bondi1944} accretion. This was because, if the initial orbital separation was too close during the rapid expansion of the AGB radius during the TP-AGB phase, a common-envelope \citep{Paczynski1976} would result. On a timescale of 100--1000\,yrs, this common-envelope would drastically shrink the orbital period and result in the ejection of the TP-AGB envelope. This envelope ejection would truncate the thermal pulses, and if there had not already been sufficient carbon enhancement, the proto-dC would remain a normal O-rich main-sequence star. 

However, there are numerous cases of short-period dCs that have been found. \citet{Miszalski2013} found that the central star of the ``Necklace'' planetary nebula is in a binary with a dC companion, with an orbital period of 1.16\,d \citep{Corradi2011}. Here, we see direct evidence of the common-envelope \added{phase} and the expelled envelope in the form of the nebula. \citet{Margon2018} found a dC without these clear signs of a common-envelope \added{phase} to have a period of 2.92\,d using multi-epoch photometric surveys. 

\citet{Roulston2021a} just published 34 new dC periods, and \citet{Whitehouse2021} found 5 new dC periods. Remarkably, 95\% of this combined sample of dCs have P$< 10$\,d (with the shortest being from \citet{Roulston2021a} with a period of only 3.2\,hr). \added{We note that both the \citet{Roulston2021a} and \citet{Whitehouse2021} samples used photometric surveys that are optimized for finding the types of short periods seen in their dC samples. Known dC periods range up $\sim$11\,yrs, which would require long term photometric, spectroscopic or astrometric observations to detect and confirm.} However, it is clear that dCs can go through a common-envelope \added{phase}, and possibly even a significant fraction do so.

\subsection{dC Rotation and Activity}
\begin{figure*}
\gridline{\fig{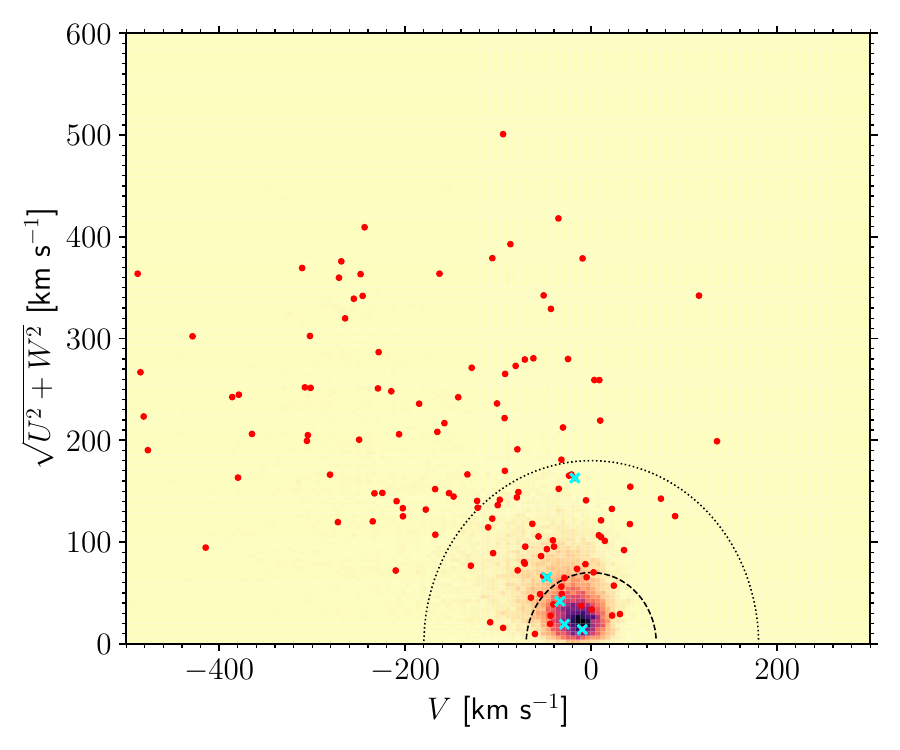}{0.5\textwidth}{(a)} 
          \fig{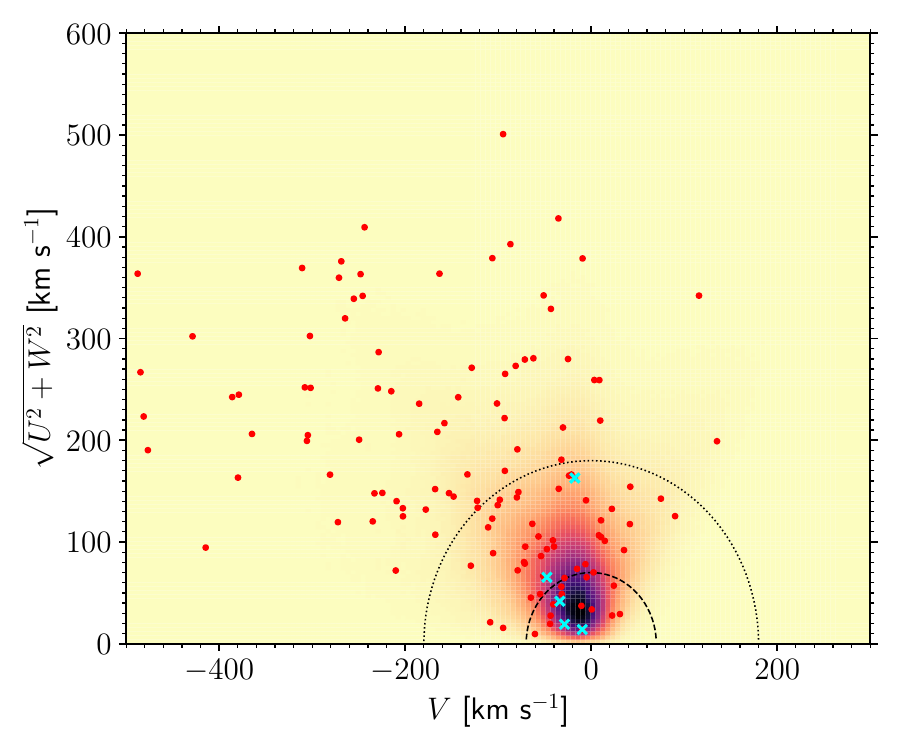}{0.5\textwidth}{(b)}}
\caption{Toomre diagram for dCs showing the UVW kinematics with the dCs shown as red scatter points (the cyan points are dCs with detected periodicity from \citealt{Roulston2021a}). The radial velocity for each dC was measured from the H$\alpha$ line \added{from their publicly available spectra, either LAMOST or SDSS}. Distances and proper motions were taken from \Gaia\ EDR3. The marked dashed and dotted lines represent, respectively, the divisions between thin and thick disks, and thick disk and halo kinematics \citep[$V_{tot.}$ = 70, 180 km s$^{-1}$;][]{Bensby2005, Reddy2006, Nissen2009}. A sample of SDSS K and M dwarfs are shown as the background heatmap, with darker colors showing regions with a higher density of K and M dwarfs. These normal (C/O $< 1$) stars are more representative of the thin and thick disk. \replaced{dCs must be in binary systems, and likely short period binary systems.}{As dCs must be in binaries, and some are known to be short period binaries, their radial velocities may be inflated by their orbital motions.} We therefore also inflate the K and M dwarfs UVW space velocities with similar velocities using the RV models of \citet{Roulston2019}. The left panel (a) shows the uninflated UVW space velocities for the K and M dwarfs, while the right panel (b) shows the inflated UVW velocities. Even when the K and M dwarfs are inflated with larger velocities to match the dCs, they do not match the same UVW kinematics as the dCs, pointing to dCs truly originating from thick disk or halo populations.}
\label{fig:toomre}
\end{figure*}

Main-sequence stars are known to ``spin down'' as their rotation rates, dynamo strengths, and associated activity decrease with age (e.g., \citealt{Kraft1967, Skumanich1972}). As dCs are thought to be from older thick disk and halo populations \citep{Green2013, Farihi2018}, they may be expected to exhibit slower rotation rates and corresponding weaker activity. To demonstrate dCs' likely population and hence age, we show a Toomre diagram \citep{Carney1988} for a sample of dCs in Figure \ref{fig:toomre}. We used the \citet{Green2013} SDSS sample of C stars, matching to \Gaia\ EDR3 \citep{GaiaEDR3}. We selected only those stars with: (1) parallax $\varpi / \varpi_{\rm{err}} > 5$ (2) proper motion signal-to-noise $> 5$ in both right ascension and declination and (3) absolute $M_G > 5$. We measured the dC radial velocities from the H$\alpha$ line, and then used the \Gaia EDR3 distance and proper motions to calculate the space velocities $U, V, W$. We mark the transitions between thin and thick disk, and thick disk and halo kinematics. As seen in the figure, the majority of dCs show kinematics consistent with either thick disk or halo populations.

However, dCs' activity may not correlate simply with age, because they are not single stars and therefore do not evolve independently. Indeed, dCs reveal a population of binary systems in which interaction and mass transfer can be confirmed by simple inspection of their signature optical spectra. 


\citet{Jeffries1996} showed that a slow (10--20\,km\,s$^{-1}$) AGB wind can spin up a low-mass companion to short ($\la 10$\,hr) rotation periods. If dC stars gain most of their carbon-rich mass through wind-Roche Lobe Overflow \citep[WRLOF;][]{Mohamed2007}, which focuses the wind into the orbital plane, it is possible that this may cause dCs to spin up to even shorter periods. Rapid rotation in stars with convective envelopes drives a magnetic dynamo, so this spin-up rejuvenation may result in enhanced chromospheric and coronal activity (e.g., \citealt{Kosovichev2013}), which normally yield observable H$\alpha$ and/or X-ray emission. Since M dwarfs show activity lifetimes of $\sim 1$--$5$\,Gyr \citep{West2008}, dCs may remain active after mass transfer for similar timescales. Additionally, \citet{Matrozis2017} modeled the maximum amount of mass the progenitors of the better studied Ba and CEMP-s stars can accrete before reaching critical rotation. They found that in order for these stars, and by similarity dCs, to accrete enough material to change their surface abundances there must be angular momentum loss from the freshly spun-up accretor. They suggest one possible method of angular momentum loss is through enhanced magnetic braking from the increased differential rotation of the accretor envelope.

\citet{Green2019} thus aimed to study the activity and rejuvenation of dCs using \Chandra. Their sample was constructed to observe the dCs that were most likely to be detected based on optical spectroscopy, i.e., those with either H$\alpha$ emission or showing a composite dC+WD spectrum. They detected all six members of their observed sample; however, their sources lacked enough counts to robustly fit a model to the source spectrum. Nonetheless, they fit two models appropriate for coronally active stars, with differing plasma temperatures of 2\,MK and 10\,MK. \citet{Green2019} found that when assuming the lower 2\,MK plasma temperature, dCs populate the saturated regime where $\log{(\LxLbol)} \sim -3.3$ \citep[e.g.,][]{Wright2011}, indicating short rotation periods. However, with the higher 10MK plasma temperatures, only half of the dCs remain in the saturated regime, with periods weakly constrained to $< 20$\,d. While there were no rotation periods available in \citet{Green2019}, they argued that their saturated X-ray activity indicated rapid rotation rates that were indicative of dC spin up from mass transfer. 

\citet{Green2019} end their discussion with the caveat that their sample is not representative of dCs in general since they explicitly observed those dCs with optical signs of activity. They argue that observations of a sample of the closest dCs, without requiring signs of activity, is critical to understanding the dC rotation-activity relationship. Here, we have targeted such a sample, observing the five closest known dCs.

\section{Sample Selection}\label{sec:sample}


We compiled our parent sample of dCs from the current literature. The largest contributor (747 dCs, 79\%) is the \citet{Green2013} sample of carbon stars from the Sloan Digital Sky Survey \citep[SDSS;][]{SDSS_1-2}. We also selected a smaller number of dCs from \citet{Si2014}, who found 96 new dCs using a label propagation algorithm from SDSS DR8, and \citet{Li2018} who selected carbon stars from the Large Sky Area Multi-Object Fiber Spectroscopic Telescope survey \citep[LAMOST;][]{LAMOST} using a machine learning approach. Our resulting final \added{parent} sample consists of 944 dCs, where we ensured that each is indeed a dC by verifying that each C star had M$_G > 5$\,mag from \Gaia\ Early Data Release 3 \citep[\Gaia\ EDR3;][]{GaiaEDR3} while having a significant parallax of $\varpi / \varpi_{\rm{err}} > 3$.  \added{We expect to publish this large parent sample, along with detailed SED fit parameters in an upcoming paper.}

\begin{figure}
\centering
\epsscale{1.15}
\plotone{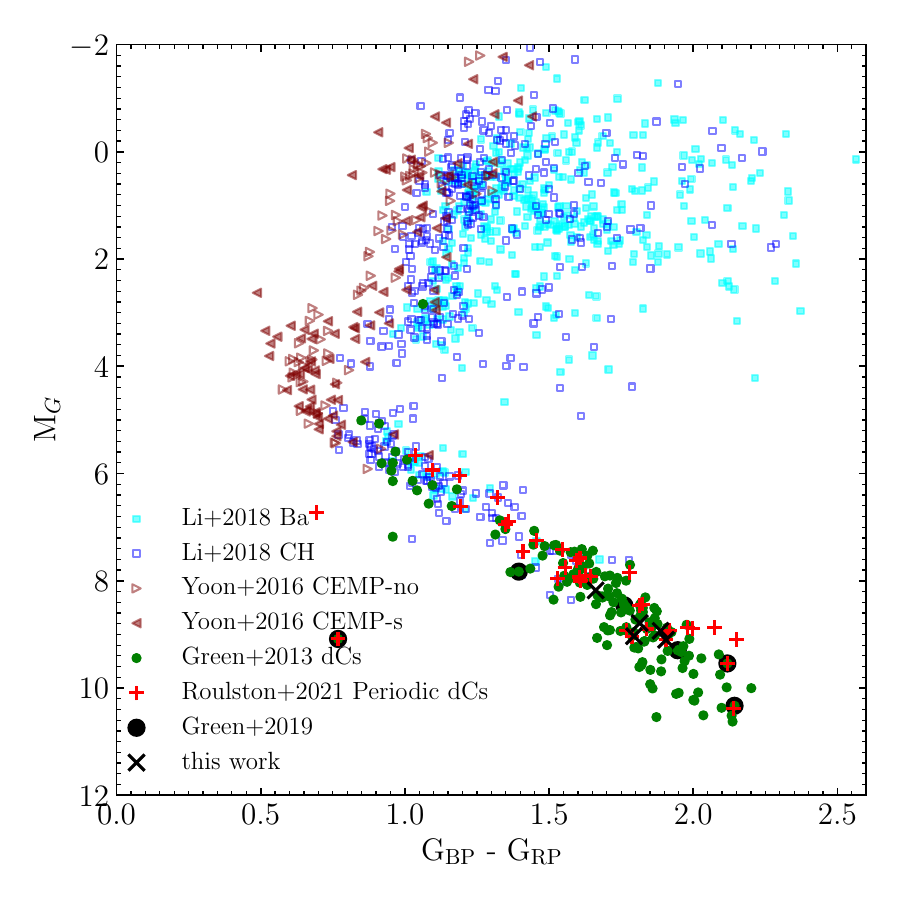}
\caption{Updated color-magnitude diagram from \citet{Green2019}. While we only analyze main sequence stars in this paper, we include the C-rich subgiants and giants in this plot merely to illustrate that they may be the evolved descendants of dC stars, which all together present a C-rich HR diagram.  Further studies comparing e.g., the binary orbital properties, kinematics and abundances of these samples can confirm this hypothesis. We show only those carbon-enhanced stars with $\varpi/\sigma_\varpi > 10$ from  \Gaia\ EDR3 \citep{GaiaEDR3}. The majority of dCs come from the SDSS samples of \citet{Green2013}, \citet{Si2014}, and the LAMOST sample from \citet{Li2018}. We show the more evolved carbon-enhanced stars (i.e. Ba, CH, and CEMP-s/no stars) from the  \citet{Yoon2016}  and \citet{Li2018} samples. Additionally, we show important samples of dCs: the sample of periodic dCs from \citet{Roulston2021a}, the \Chandra\ sample from \citet{Green2019} and the sample from this work, with the later two populating the low-luminosity red tail of this sequence.}
\label{fig:CMD}
\end{figure}

We then selected the nearest five dCs to make our final \added{\Chandra} sample. Our selected sample can be found in Table \ref{tab:distLbol} with their corresponding  \Gaia\ EDR3 properties. In addition, we have estimated the bolometric luminosities for each selected dC in the same way as in \citet{Green2019} by using a spectral energy distribution (SED) for each dC and the \texttt{sedkit} Python package \citep{Filippazzo2015}. Figure \ref{fig:CMD} shows a color-magnitude diagram for a sample of chemically-enhanced stars, including dCs. From this, we can see that the five dCs in this sample are all clearly dwarfs having M$_G > 8$. \added{Additionally, we examined the available spectra for the five dCs selected here, all of which show strong C$_2$ and CN bands.}

The reddest stars in this CMD correspond to the lowest mass dCs. These dCs will likely have accreted a substantial fraction (see \citet{Miszalski2013} and the discussion at the end of this paper) of their current mass, and may have even been brown dwarfs before the onset of the accretion that turned them into dCs (see \citet{Majidi2021} for a discussion on a similar topic).

\begin{deluxetable*}{llcccccc}
\tablecaption{Parallaxes, Distances and Bolometric Luminosities}
\tablecolumns{5}
\tablewidth{0pt}
\tablehead{
\colhead{Object} &
\colhead{} &
\colhead{R.A.} &
\colhead{Decl.} &
\colhead{$\varpi$} &
\colhead{Distance} &
\colhead{M$_G$} &
\colhead{log (\Lbol/$L_\odot$)}
\\
 && \colhead{(J2016.0)} & \colhead{(J2016.0)} & \colhead{[mas]} & [pc] & [mag] & 
}
\startdata
\object[LSPM J0435+3401]{LSPM J0435+3401} &(J0435)                       & 04h35m26.31s & +34d01m35.54s & 7.819 $\pm$ 0.019 & 127.19 $\pm$ 0.21 & 8.18 & -1.27 $\pm$ 0.01 \\
\object[LAMOST J054640.48+351014.0]{LAMOST J054640.48+351014.0} &(J0546) & 05h46m40.51s & +35d10m13.23s & 6.126 $\pm$ 0.035 & 162.05 $\pm$ 0.62 & 9.03 & -1.54 $\pm$ 0.01 \\
\object[HE 1205-0417]{HE 1205-0417} & (HE1205)                          & 12h07m51.75s & -04d34m41.55s & 6.52 $\pm$ 0.12   & 153 $\pm$ 2       & 8.90 & -1.43 $\pm$ 0.01 \\
\object[LAMOST J124055.15+485114.2]{LAMOST J124055.15+485114.2} &(J1240) & 12h40m55.17s & +48d51m14.14s & 5.57 $\pm$ 0.15   & 179 $\pm$ 3       & 8.93 & -1.55 $\pm$ 0.02 \\
\object[SBSS 1310+561]{SBSS 1310+561} &  (SBSS1310)                         & 13h12m42.27s & +55d55m54.84s & 9.538 $\pm$ 0.023 & 104.48 $\pm$ 0.17 & 9.10 & -1.51 $\pm$ 0.01 \\
\enddata
\tablecomments{Selected dCs in our sample. For each, we list the coordinate positions and parallaxes from  \Gaia\ EDR3 \citep{GaiaEDR3}. We take the distance for each dC from \citet{GaiaEDR3_dist}, which were used to calculate the absolute Gaia G band magnitude. Additionally, for each dC, we include the bolometric luminosity as calculated from our SED fits. Here, we adopt the solar bolometric luminosity of $\log{\left(L_\sun / \textrm{erg s}^{-1}\right)} = 33.583$.}
\end{deluxetable*}
\label{tab:distLbol}

\section{X-Ray Observations and Analysis}\label{sec:xrays}

All of our \Chandra\ observations were taken with ACIS-S using the S3 chip (backside illuminated CCD) between 2019 September and \replaced{2020}{2021} December (\Chandra\ proposals 21200072 and 22200008; PI: P. Green). Exposure times, proposed based on optical magnitudes, ranged from 9.83\,ks to 36.78\,ks, with one dC (\object[LSPM J0435+3401]{LSPM J0435+3401}) having multiple exposures (ObsIDs). None of our observations use a grating and all were taken in VFAINT ACIS mode; observation details can be found in Table \ref{tab:cxc_props}.

We reprocessed the \Chandra\ event lists for all ObsIDs with the CIAO (ver 4.13.0) \texttt{chandra\_repro} script and CALDB (4.9.5). This reprocessing ensures we have used the most recent calibrations for our data, including corrections for afterglows, bad pixels, charge transfer inefficiency, and time-dependent gain corrections. We searched each ObsID for each target near the expected  \Gaia\ EDR3 position, detecting sources for two of our five targets, J0435 and SDSS1310. We detected J0435 in all seven ObsIDs, with all detections being within 1.1\arcsec\ of the expected  \Gaia\ EDR3 position. SDSS1310 was detected within 0.3\arcsec\ of the  \Gaia\ EDR3 position. A detection was counted if there were multiple neighboring pixels, within 5\arcsec of the expected source, that had at least one X-ray count.

We estimated source properties in the $0.3$--$8.0$\,keV energy range using the CIAO \texttt{srcflux} tool for the two dCs with detections. For both detections, we used a circular aperture of 5\arcsec\ for the source region and an annular aperture with an inner radius of 5\arcsec\ and outer radius of 15\arcsec\ for the background region. Both of these regions were centered on the detected position of the dC. We accounted for Milky Way dust extinction by using the 3-dimensional Bayestar17 \citep{Bayestar2017} dust maps from the \texttt{dustmaps} Python package \citep{dustmaps}; resulting line of sight column density ($N_H$) values are listed in Table \ref{tab:cxc_props}. 

\begin{deluxetable*}{cccccccccc}
\tablecaption{{\em Chandra} X-ray Observations}
\tablecolumns{6}
\tablewidth{0pt}
\tablehead{
\colhead{Object} & \colhead{ObsID} & \colhead{Obs-Date} & \colhead{Exposure}  & \colhead{Net Count Rate} &
\colhead{$N_{\rm H}$} & \colhead{$T_{\rm X}$} & \colhead{$F_{\rm X,obs}$} &  \colhead{$F_{\rm X}$} &  \colhead{$L_{\rm X}$}
\\
 & &  & \colhead{[ks]}& \colhead{[cnt ks$^{-1}$]}  & \colhead{[$10^{19} $ cm$^{-2}$]}  & \colhead{[MK]} & \multicolumn{2}{c}{[$10^{-15}$ erg~cm$^{-2}$~s$^{-1}$]} & \colhead{[$10^{28}$ erg s$^{-1}$]}
}
\startdata
J0435    & \dataset[22298]{https://doi.org/10.25574/22298} & 2019-09-15 & 10.12  & $2.61^{+0.33}_{-0.27}$ & $6.3^{+0.7}_{-1.1}$ & 2  & $230 ^{+57 }_{-38 }$  & $246 ^{+61 }_{-41 }$  & $48^{+12}_{-8.0}$ \\
\ldots   & \ldots                                          & \ldots     & \ldots & \ldots                 & \ldots              & 10 & $31.6^{+7.8}_{-5.2}$  & $32.3^{+8.0}_{-5.3}$  & $6.3 ^{+1.6 }_{-1.0}$ \\
\ldots   & \dataset[23376]{https://doi.org/10.25574/23376} & 2020-12-15 & 19.81  & $2.39^{+0.22}_{-0.19}$ & \ldots              & 2  & $319 ^{+56 }_{-42 }$  & $341 ^{+60 }_{-45 }$  & $66   ^{+12}_{-9.0}$ \\
\ldots   & \ldots                                          & \ldots     & \ldots & \ldots                 & \ldots              & 10 & $36.0^{+6.4}_{-4.7}$  & $36.8^{+6.5}_{-4.8}$  & $7.1 ^{+1.3  }_{-1.0}$ \\
\ldots   & \dataset[24305]{https://doi.org/10.25574/24305} & 2020-12-20 & 9.83   & $2.23^{+0.91}_{-0.25}$ & \ldots              & 2  & $299 ^{+83 }_{-53 }$  & $319 ^{+89 }_{-57 }$  & $62^{+17}_{-11}$ \\
\ldots   & \ldots                                          & \ldots     & \ldots & \ldots                 & \ldots              & 10 & $33.7^{+9.3}_{-6.0}$  & $34.5^{+9.6}_{-6.2}$  & $6.7 ^{+1.9 }_{-1.2}$ \\
\ldots   & \dataset[24306]{https://doi.org/10.25574/24306} & 2021-12-20 & 11.42  & $2.50^{+0.55}_{-0.45}$ & \ldots              & 2  & $428 ^{+185 }_{-125 }$& $458 ^{+199 }_{-134 }$  & $88^{+38}_{-26}$ \\
\ldots   & \ldots                                          & \ldots     & \ldots & \ldots                 & \ldots              & 10 & $42.3^{+18.0}_{-12.0}$& $43.3^{+19.0}_{-13.0}$  & $8.4 ^{+3.6 }_{-2.5}$ \\
\ldots   & \dataset[24893]{https://doi.org/10.25574/24893} & 2020-12-21 & 17.83  & $2.77^{+0.25}_{-0.21}$ & \ldots              & 2  & $371 ^{+64 }_{-48 }$  & $397 ^{+68 }_{-51 }$  & $77^{+13}_{-10}$ \\
\ldots   & \ldots                                          & \ldots     & \ldots & \ldots                 & \ldots              & 10 & $41.8^{+7.2}_{-5.4}$  & $42.7^{+7.4}_{-5.5}$  & $8.3 ^{+1.4}_{-1.1}$ \\
\ldots   & \dataset[24896]{https://doi.org/10.25574/24896} & 2020-12-20 & 19.81  & $3.17^{+0.23}_{-0.23}$ & \ldots              & 2  & $424 ^{+56 }_{-56 }$  & $454 ^{+60 }_{-60 }$  & $88^{+12}_{-12}$ \\
\ldots   & \ldots                                          & \ldots     & \ldots & \ldots                 & \ldots              & 10 & $47.8^{+6.3}_{-6.3}$  & $48.9^{+6.5}_{-6.5}$  & $9.5 ^{+1.3 }_{-1.3}$ \\
\ldots   & \dataset[26242]{https://doi.org/10.25574/26242} & 2021-12-20 & 17.23  & $2.76^{+0.46}_{-0.40}$ & \ldots              & 2  & $472^{+83 }_{-61 }$   & $505 ^{+89 }_{-65 }$  & $98^{+17}_{-13}$ \\
\ldots   & \ldots                                          & \ldots     & \ldots & \ldots                 & \ldots              & 10 & $46.6^{+8.2}_{-6.0}$  & $47.7^{+8.4}_{-6.2}$  & $9.2 ^{+1.6 }_{-1.2}$ \\
J0546    & \dataset[22299]{https://doi.org/10.25574/22299} & 2020-01-09 & 34.80  & $< 0.20$               & $7.5^{+1.5}_{-0.6}$ & 2  & $< 18$                & $< 24$                & $< 7.2 $ \\
\ldots   & \ldots                                          & \ldots     & \ldots & \ldots                 & \ldots              & 10 & $< 2.5$               & $< 2.8$               & $< 0.9 $ \\
HE1205   & \dataset[22300]{https://doi.org/10.25574/22300} & 2020-04-04 & 24.23  & $< 0.44$               & $8.4^{+1.3}_{-1.2}$ & 2  & $< 47$                & $< 51$                & $< 14$ \\
\ldots   & \ldots                                          & \ldots     & \ldots & \ldots                 & \ldots              & 10 & $< 5.9$               & $< 6.1$               & $< 1.7 $ \\
J1240    & \dataset[22301]{https://doi.org/10.25574/22301} & 2019-09-19 & 36.78  & $< 0.34$               & $6.3^{+0.8}_{-1.2}$ & 2  & $< 31$                & $< 32$                & $< 10$ \\
\ldots   & \ldots                                          & \ldots     & \ldots & \ldots                 & \ldots              & 10 & $< 4.2$               & $< 4.3$               & $< 1.3 $ \\
SBSS1310 & \dataset[22302]{https://doi.org/10.25574/22302} & 2020-01-25 & 14.90  & $1.98^{+0.24}_{-0.19}$ & $3.0^{+1.8}_{-1.7}$ & 2  & $203 ^{+47 }_{-32  }$ & $210 ^{+49 }_{-33  }$ & $40.7^{+9.5}_{-6.4}$ \\
\ldots   & \ldots                                          & \ldots     & \ldots & \ldots                 & \ldots              & 10 & $25.9^{+6.0}_{-4.1}$  & $26.2^{+9.2}_{-4.1}$  & $5.1  ^{+1.2 }_{-0.8}$ \\
\enddata
\tablecomments{ The ObsID, date, and exposure time are listed for each individual observation. For the dCs with detections (J0435, SBSS1310), the net count rate is shown. For each detected exposure, we use the column density (N$_{\rm H}$) from \cite{Bayestar2019} and assume two different plasma temperatures of $2$\,MK and $10$\,MK. For each assumed plasma temperature, the observed source flux, unabsorbed source flux,  and source luminosity are calculated in the $0.3$--$8.0$\,keV range. The 1$\sigma$ errors for each property are shown. For dCs without a detection, the 3$\sigma$ upper limits are shown assuming the same set of plasma temperatures.}
\end{deluxetable*}
\label{tab:cxc_props}

Table \ref{tab:cxc_props} lists the ObsIDs for each dC, the observation properties, and the X-ray source properties for each dC. The calculated net count rate for each ObsID is given, with the $1\sigma$ upper and lower error limits. For the three dCs with no detections, the $3\sigma$ upper limits are given. Following \citet{Green2019}, we derived two X-ray flux estimates, using both a 2\,MK and a 10\,MK optically thin plasma \citep[APEC;][]{Smith2001} with absorption modeled using WABS \citep{Morrison1983}. We list the $1\sigma$ upper and lower limits for the two dCs with detections for the net count rate, observed flux, unabsorbed flux, and X-ray luminosity. For the three dCs without detections, we list the $3\sigma$ upper limits.

\subsection{Individual Spectral Fits}\label{sec:indiv_fits}

For the dCs SBSS1310 and J0435, we have also fit individual spectral models for each observation. These individual fits use the same APEC and WABS models as before, but now fitting the plasma temperature, column density, and a normalization as free parameters.

\subsubsection{SBSS1310}\label{sec:SBSS1310_indfit}

The best fitting model for SBSS1310 consists of a 12.1\,MK plasma temperature with a low column density ($3\sigma$ upper limit of N$_{\rm H} = 9.6\times10^{21}$\,cm$^{-2}$). This column density is consistent with the negligible Bayestar17 expected line of sight column density of N$_{\rm H} = 3\times10^{19}$\,cm$^{-2}$. 

\begin{figure*}
\gridline{\fig{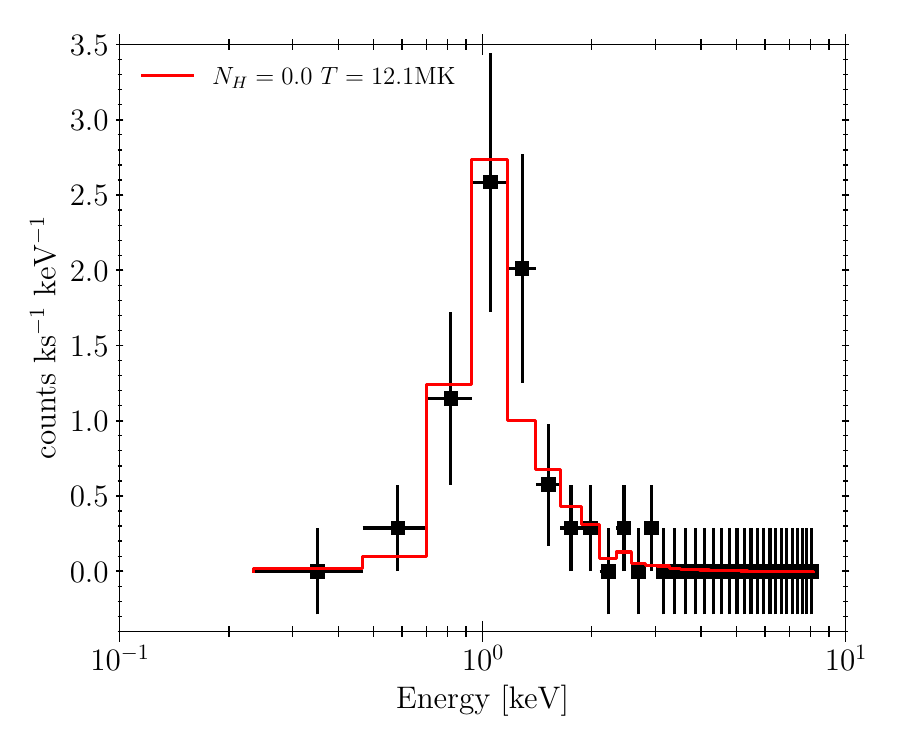}{0.5\textwidth}{(a)}
          \fig{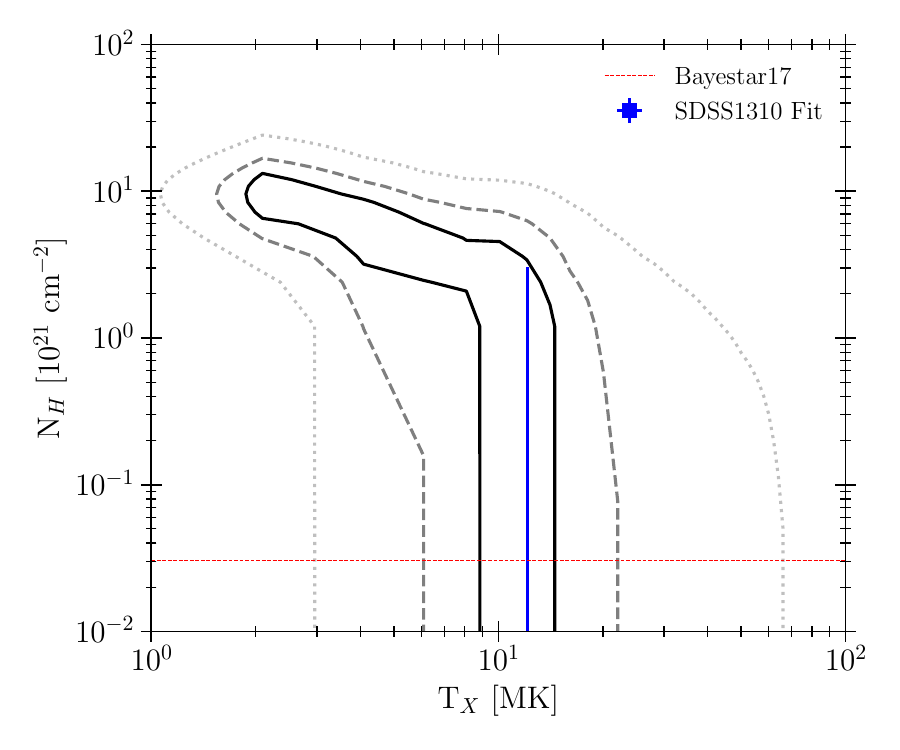}{0.5\textwidth}{(b)}}
\caption{Best fitting spectral model for SBSS13010. The left panel, (a), shows the observed source spectrum in black \deleted{scatter points} with the associated errors. The best fitting model is shown as the solid red line. The right panel, (b), shows the contours for the plasma temperature and column density, with the  $1\sigma$, $2\sigma$, $3\sigma$ contours shown as the solid, dashed and dotted lines respectively. \added{These contours were calculated by varying the plasma temperature and column density over a grid and calculating the confidence with a $\chi^2$ statistic using the Sherpa package \citep{Sherpa}.} The best fit parameters are shown as the blue marker, with the $1\sigma$ errors. The Bayestar17 expected line of sight column density is shown as the red dashed line for reference.}
\label{fig:SBSS1310_fits}
\end{figure*}

Figure \ref{fig:SBSS1310_fits} shows the best fit spectral model for SBSS1310. The left panel shows the observed source spectrum with the associated errors and the best fitting model. The right panel shows the error contours for the fit parameters, with the best fit parameters shown as the blue marker. The Bayestar17 expected line of sight column density is shown as the red dashed line for reference. The resulting model is consistent with the expected negligible column density, and with the assumed 10\,MK plasma temperature from \citep{Green2019}. \citet{Green2019} assumed this 10\,MK plasma temperature based on X-ray-selected stellar samples observed with Chandra from, e.g., the COSMOS survey \citep{Wright2010}.
\subsubsection{J0435}\label{sec:J0435_indfit}

\replaced{We carried out five}{Our requested 100\,ksec was split by \Chandra\ mission planners into seven} separate observations of J0435. As with SBSS1310, we fit each observation with a spectral model with plasma temperature and column density as free parameters. For each ObsID, the resulting plasma temperature and column density are listed in Table \ref{tab:J0435_singleFits} with their $1\sigma$ errors. All of the fits result in a column density three orders of magnitude higher than the expected line of sight column density from Bayestar17 dust map (N$_{\rm H} = 6.3\times10^{19}$\,cm$^{-2}$), \added{which suggests the presence of material around J0435 (see Section \ref{sec:J0435_SED})}. The fit plasma temperatures are consistent with the higher assumed values of the 10\,MK models. Figure \ref{fig:J0435_NHvsTx} shows the resulting best fit parameters for J0435. The colored scatter points show the best fitting parameters for the individual fits, with their errors.  

\begin{deluxetable}{ccc}
\tablecaption{J0435 Individual Model Fits}
\tablecolumns{6}
\tablewidth{0pt}
\tablehead{
\colhead{ObsID} &
\colhead{$N_{\rm H,fit}$} & \colhead{$T_{\rm X,fit}$} 
\\
  &  \colhead{[$10^{22}$ cm$^{-2}$]} & \colhead{[MK]} 
}
\startdata
\dataset[22298]{https://doi.org/10.25574/22298} & $1.43^{+0.33}_{-0.32}$ & $33^{+11}_{-5.0}$ \\
\dataset[23376]{https://doi.org/10.25574/23376} & $2.07^{+0.25}_{-0.13}$ & $9.1 ^{+1.3}_{-1.3}$   \\
\dataset[24305]{https://doi.org/10.25574/24305} & $1.26^{+0.30}_{-0.18}$ & $10.1^{+1.8}_{-1.3 }$  \\
\dataset[24306]{https://doi.org/10.25574/24306} & $3.41^{+1.44}_{-0.44}$ & $6.7^{+1.5}_{-1.4 }$  \\
\dataset[24893]{https://doi.org/10.25574/24893} & $1.48^{+0.14}_{-0.32}$ & $13.9^{+1.8}_{-1.2 }$  \\
\dataset[24896]{https://doi.org/10.25574/24896} & $1.48^{+0.22}_{-0.28}$ & $13.6^{+1.4}_{-1.4 }$  \\
\dataset[26242]{https://doi.org/10.25574/26242} & $1.77^{+0.87}_{-0.63}$ & $18.4^{+2.6}_{-1.5 }$  \\
\enddata
\tablecomments{Individual model fits for each ObsID of J0435. Each model uses the same model, but is fit independently. The $1\sigma$ errors for each fit parameter are shown.}
\label{tab:J0435_singleFits}
\end{deluxetable}

In addition to the individual fits, we simultaneously fit all \replaced{five}{seven} observations with one model. For this fit, we used the same APEC and WABS models as before, but all \replaced{five}{seven} observations are fit with the same plasma temperature and column density, allowing only the normalization to vary between each observation. 

\begin{figure*}
\centering
\plotone{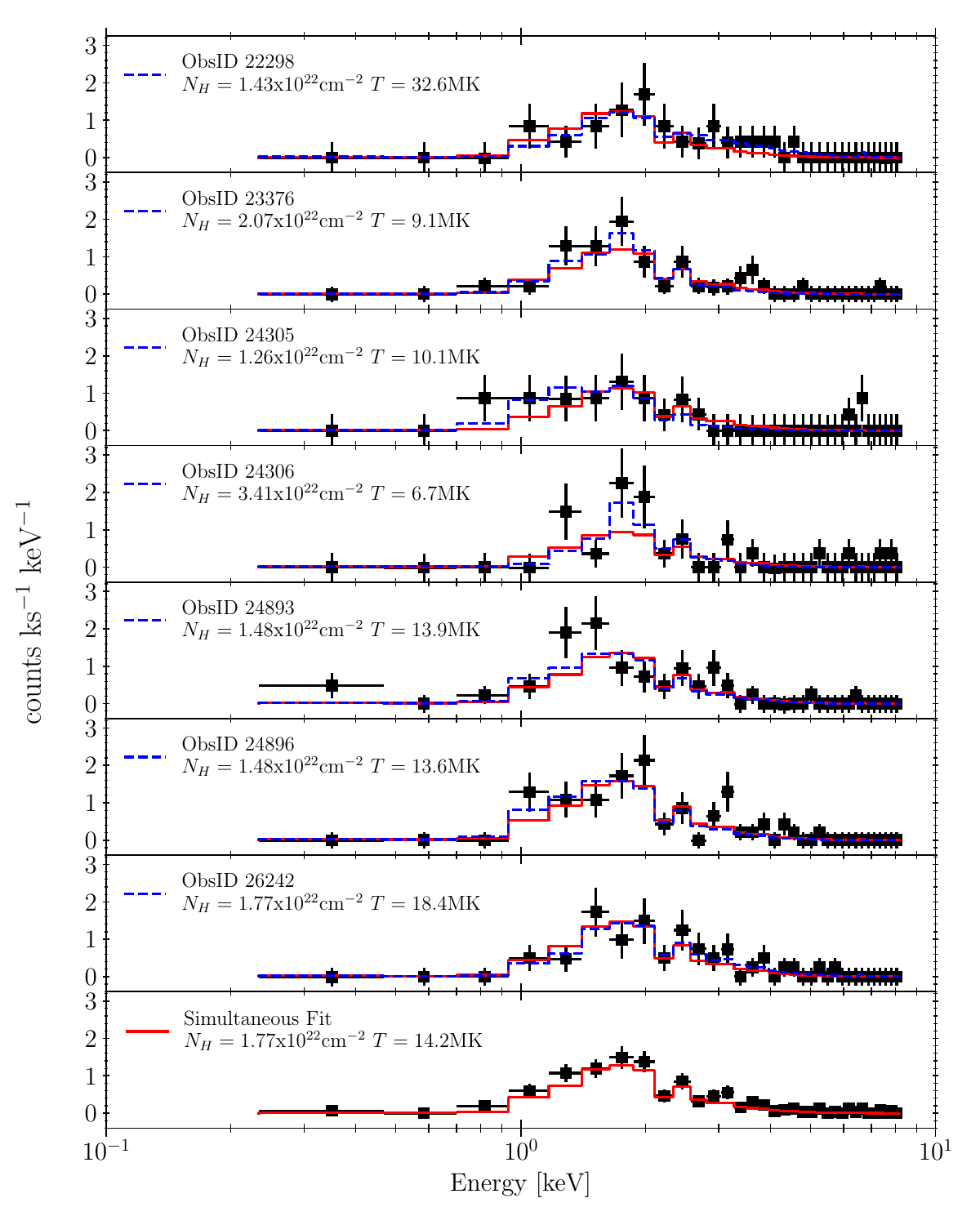}
\caption{\Chandra\ count rate spectrum in the range $0.3$--$8.0$\,keV for J0435. Each of the seven individual ObsIDs is shown, respectively, in the upper seven panels. Each of the individual observations is fit separately with the same model. The resulting fit for each is shown as the blue dashed line in each panel, with the fit parameters shown in the panel. The combined spectrum of all seven observations is shown in the bottom (last) panel. We fit all seven observations simultaneously, forcing the same column density and plasma temperature, but leave the model normalization free to be fit for each observation. The resulting simultaneous fit is shown as the solid red line in each panel.}
\label{fig:J0435_all_fit}
\end{figure*}

The simultaneous fit results in a column density of N$_{\rm H}=1.77\pm0.30\times10^{22}$\,cm$^{-2}$ and a plasma temperature of T$_{\rm X} = 14.2\pm2.9$\,MK. Figure \ref{fig:J0435_all_fit} shows both the individual and simultaneous fits for J0435. For each observation, the observed source spectrum is shown, with the best fitting individual model shown as a dashed blue line. The best fit simultaneous model is shown as the red solid line. The bottom panel shows the combined source spectrum, with the total simultaneous fit model.

\begin{figure}
\centering
\epsscale{1.15}
\plotone{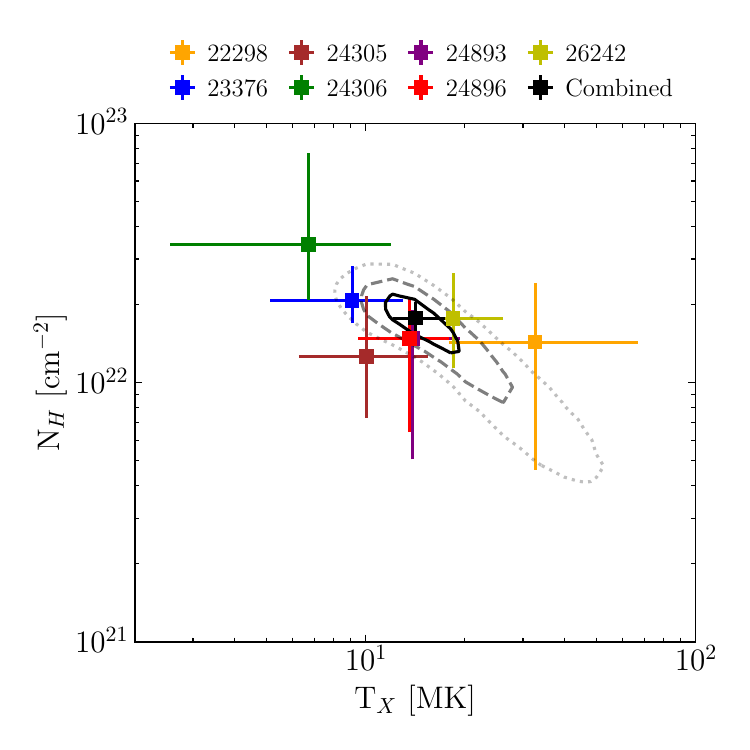}
\caption{Fit values of the model column density (N$_{\rm H}$) and plasma temperature (T$_{\rm X}$) for the dC J0435. The individual observations are shown as the color scatter points with their respective 1$\sigma$ errors. The red scatter point is the best fit for the combined set of all seven observations. The black contours represent the 1, 2, and 3$\sigma$ regions for the combined fit. The expected column density at the distance of J0435 from the 3D optical/IR dust maps of \citet{Bayestar2019} (N$_{\rm H} = 6.3\times10^{19}$\,cm$^{-2}$) is more than two orders of magnitude lower than the values shown here from X-ray fitting. }
\label{fig:J0435_NHvsTx}
\end{figure}

\begin{figure}
\centering
\epsscale{1.15}
\plotone{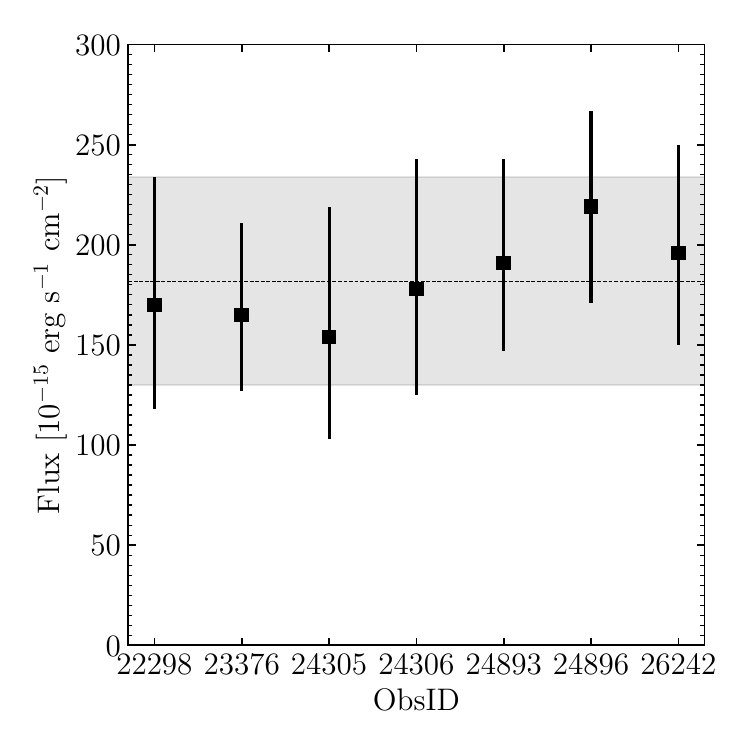}
\caption{Unabsorbed model flux for each ObsID of the dC J0435. For each ObsID, the flux is calculated using the combined fit model parameters (see Figures \ref{fig:J0435_all_fit} and \ref{fig:J0435_NHvsTx}). The dashed black line is the mean of the seven observations, with the grey shaded region showing the mean error of the observations. We find no signs of variability in the source flux of J0435. Note that all the ObsIDs for J0435 span some 28 months, but several are clustered within a few days so we plot by ObsID here, which does not map to MJDs (listed in Table\,\ref{tab:cxc_props}). }
\label{fig:J0435_flux_var}
\end{figure}

\begin{deluxetable}{cccc}
\tablecaption{J0435 Fit Source Flux}
\tablecolumns{6}
\tablewidth{0pt}
\tablehead{
\colhead{ObsID} &
\colhead{$F_{\rm X,obs}$} &  \colhead{$F_{\rm X}$} &  \colhead{$L_{\rm X}$}
\\
  &  \multicolumn{2}{c}{[$10^{-15}$ erg~cm$^{-2}$~s$^{-1}$]} & \colhead{[$10^{28}$ erg s$^{-1}$]}
}
\startdata
\dataset[22298]{https://doi.org/10.25574/22298} & $23.8^{+5.5}_{-4.4}$ & $170^{+39}_{-32}$ & $32.9^{+7.5}_{-6.1}$ \\
\dataset[23376]{https://doi.org/10.25574/23376} & $23.1^{+3.9}_{-3.3}$ & $165^{+28}_{-23}$ & $31.9^{+5.4}_{-4.5}$ \\
\dataset[24305]{https://doi.org/10.25574/24305} & $21.5^{+5.5}_{-4.3}$ & $154^{+40}_{-31}$ & $29.8^{+7.6}_{-6.0}$ \\
\dataset[24306]{https://doi.org/10.25574/24306} & $24.9^{+5.5}_{-4.5}$ & $178^{+40}_{-32}$ & $34.5^{+7.6}_{-6.2}$ \\
\dataset[24893]{https://doi.org/10.25574/24893} & $26.7^{+4.4}_{-3.7}$ & $191^{+32}_{-27}$ & $37.0^{+6.1}_{-5.2}$ \\
\dataset[24896]{https://doi.org/10.25574/24896} & $30.6^{+4.1}_{-4.1}$ & $219^{+29}_{-29}$ & $42.4^{+5.6}_{-5.6}$ \\
\dataset[26242]{https://doi.org/10.25574/26242} & $27.4^{+4.6}_{-3.9}$ & $196^{+33}_{-28}$ & $37.9^{+6.4}_{-5.4}$ \\
\enddata
\tablecomments{Combined model fits for each ObsID of J0435. For each observation, the observed source flux, unabsorbed flux, and luminosity are listed. For each value the $1\sigma$ errors are shown.}
\label{tab:J0435_simFits}
\end{deluxetable}

\subsection{X-ray Variability}\label{sec:xray_var}

Since coronal activity is by nature variable, we searched for signs of X-ray variability in the dCs observed to date. We used the CIAO implementation of the Gregory-Loredo variability algorithm \citep{Gregory1992} \texttt{glvary}. We tested all available ObsIDs for dCs in both the \citet{Green2019} sample and this work. We found that none of the dCs in \citet{Green2019} show significantly variable count rates, with all having a variability index\footnote{\url{https://cxc.cfa.harvard.edu/ciao/ahelp/glvary.html}} of 0 or 1.

From this work, we found that SBSS1310 has a variability index of 2 and is considered not variable. For J0435, the variability indices are \replaced{6, 2, 0, 0, 6}{6, 2, 0, 0, 0, 6, 0} for the ObsIDs \replaced{22298, 23376, 24305, 24893, 24896, and 26242}{22298, 23376, 24305, 24306, 24893, 24896, and 26242} respectively. The ObsIDs with variability index equal to 6 are considered definitely variable; variability index equal to 2 is considered probably not variable, and variability indexes of 0 are considered definitely not variable.

One possible explanation for this changing variability index may lie with the source of the X-ray emission in dCs, which is believed to be from coronal emission associated with rotation and magnetic activity. Stellar flares are associated with magnetic activity and magnetic reconnection \citep{Pettersen1989}, where most of the flare coronal emission is in X-rays. As dCs show X-ray emission and are thought to have rejuvenated activity, we expect dCs to flare at similar rates as active M dwarfs. As flares are transient, stochastic events, the resulting X-ray emission will also be transient and stochastic. If J0435 has an active corona, we should observe some continuous level of X-ray emission. During a flare however, the X-ray emission should increase with a lifetime of the flare. These flares could be the source of the differing levels of variability between the J0435 observations. \added{To check for flares from J0435, we searched the \textit{TESS} \citep{Ricker2015} light-curve but found no detected flare events in the full frame images. However, given that for cool stars the flare duration is of order one hour \citep{Howard2019}, the 30\,min cadence of this light curve may not resolve any flares outside of single point outliers.}



Additionally, for J0435, we used the best simultaneous fit model parameters to calculate the observed flux, unabsorbed flux, and source luminosity in the $0.3$--$8.0$\,keV range for each ObsID. These values are listed in Table \ref{tab:J0435_simFits} with their $1\sigma$ upper and lower limits. Figure \ref{fig:J0435_flux_var} shows the unabsorbed model flux for each ObsID. The $1\sigma$ error bars are shown for each, with the shaded region showing the average $1\sigma$ error across all \replaced{five}{seven} observations. Within these errors, we find no detectable variability in the source flux of J0435. 

\subsection{Rotation-Activity Relationship}\label{sec:rotation}

In main-sequence stars, X-ray emission, and often chromospheric H$\alpha$ emission, is associated with coronal activity due to magnetic activity. This activity is thought to be produced by an $\alpha\Omega$ dynamo \citep{Parker1955} which requires a differentially rotating convective envelope and a solidly rotating radiative core. However, it has been found that even late-type, fully convective stars show magnetic activity associated with rotation \citep{Wright2018}. The Rossby number $Ro = P_{rot}/\tau$ \citep{Noyes1984}, which relates the rotation period to the convective turnover time ($\tau$), has been shown to correlate with activity and saturates for rapid rotators at the level of  $\log{(\LxLbol)} \approx -3.3$ for $Ro \lesssim 0.13$ \citep{Micela1985, Wright2011}. 

In \citet{Green2019}, all six of the observed dCs were detected with \Chandra, with $\log{(\LxLbol)}$ ranging from $-4.5$ to  $-3.2$, depending on the assumed model plasma temperature. These values place the dCs in the saturated regime for stellar rotation; however, at the time, no rotation periods were known for these stars. The recent studies by \citet{Roulston2021a} and \citet{Whitehouse2021} have found many new periods for dC stars, including five of the six in the \citet{Green2019} sample. One of the detections in this work, SBSS1310, is in both  \citet{Roulston2021a} and \citet{Whitehouse2021}. However, J0435 does not have a known rotation period in the literature.

\begin{figure*}
\centering
\epsscale{1.2}
\plotone{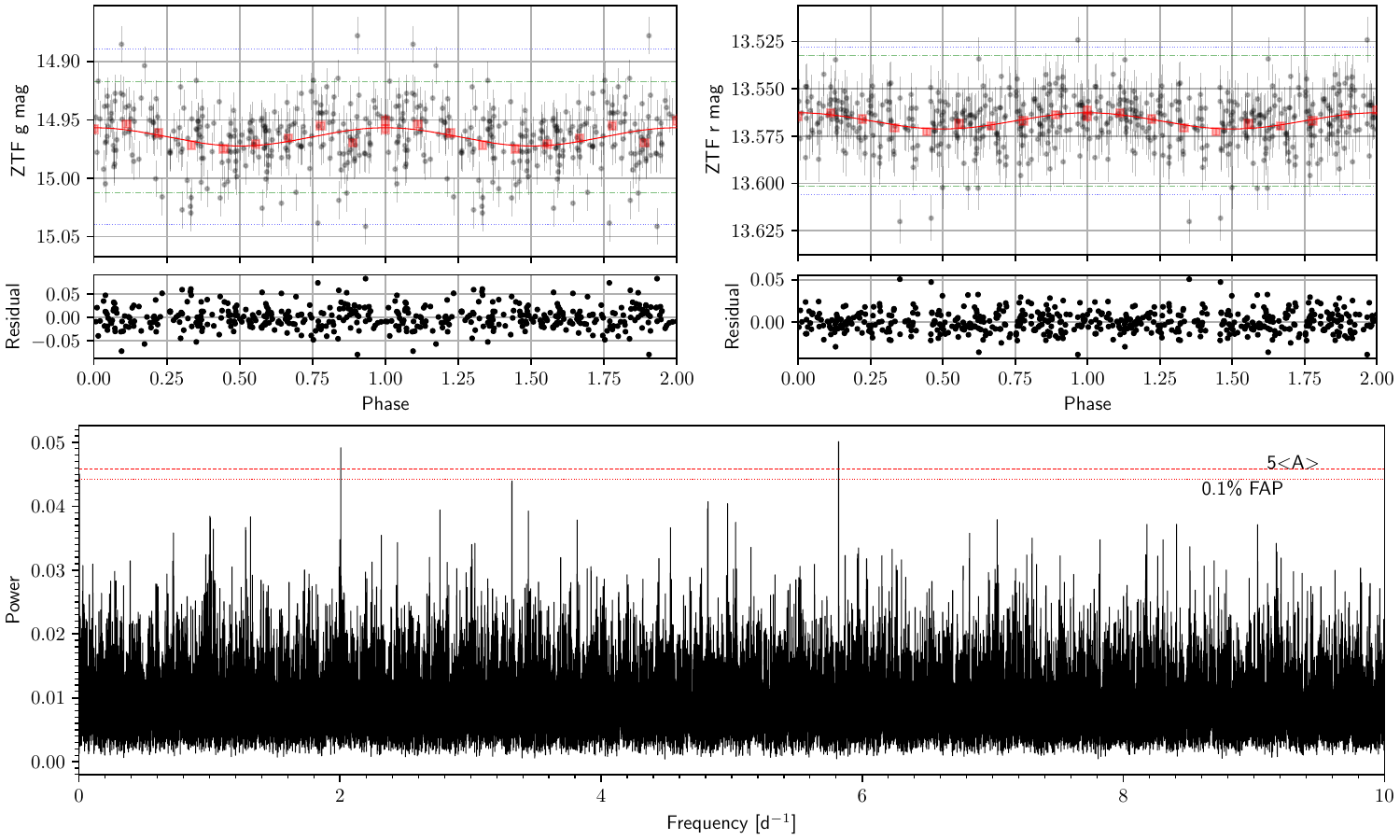}
\caption{Zwicky Transient Facility light-curves for J0435. The top panels show the light-curves in both the $g$ and $r$ bands, folded on the best found significant period of $0.1719\pm0.0016$\,d. The red line shows the best fitting model to the each phased light-curve, with residuals shown below the light-curve. The red square markers are the binned data in 10 phase bins. The multiband LS power spectrum is shown in the bottom panel, Our 5$\langle A\rangle$ or $0.1\%$\,FAP limits are marked by the dashed and dotted red horizontal lines respectively.}
\label{fig:J0435_ZTF_LC}
\end{figure*}

We searched the light-curve of J0435 in the Zwicky Transient Facility DR5 \citep[ZTF;][]{ZTF_1,ZTF_2,ZTF_3}. for periodic signals. We used similar methods as detailed in \citet{Roulston2021a}, explained briefly here. We used an outlier removal procedure to clean the raw light-curve, before searching for periodic signals using a Lomb-Scargle periodogram \citep[LS;][]{Lomb1976, Scargle1982}. While there were no significant peaks in the individual ZTF light curves, we used the multiband periodogram from \citet{astropyLS2} to search for shared variability in both the $g$ and $r$ bands. To consider a peak in the power spectrum as significant, we used the 5$\langle A\rangle$ (five times the mean power) limit as well as the $0.1\%$\,false-alarm probability (FAP) limit (see \citealt{Greiss2014} and \citealt{Hermes2015} for more details). Additionally, we required that the peak frequency must be separated by at least 0.005\,d$^{-1}$ from an observational alias, such as 0.5\,d$^{-1}$ or 0.333\,d$^{-1}$. 

The highest peak in the combined ZTF power spectrum for J0435 meets both of these requirements, so we take the period to be $0.1719\pm0.0016$\,d. We do note the caveat that this period is assumed to be both the rotation period and orbital period under the assumption that in a close binary system, we would expect a synchronized \citep{Hurley2002}, low eccentricity orbit. Figure \ref{fig:J0435_ZTF_LC} shows the ZTF light-curve for J0435 folded on the highest found significant peak. The best fitting single sinusoidal model to the data is shown for each band as the solid red line, with the residuals below. The bottom panel shows the power spectrum, and the power needed to reach our  5$\langle A\rangle$ or $0.1\%$\,FAP limits. J0435 also has a light curve in the Catalina Real-Time Transient Survey \citep{CRTS1}, but including this light curve in the multiband periodogram results in the same period and significance as the ZTF only analysis. Since the Catalina data have much larger errors, we do not include them in our multiband analysis. 

\begin{figure*}
\centering
\plotone{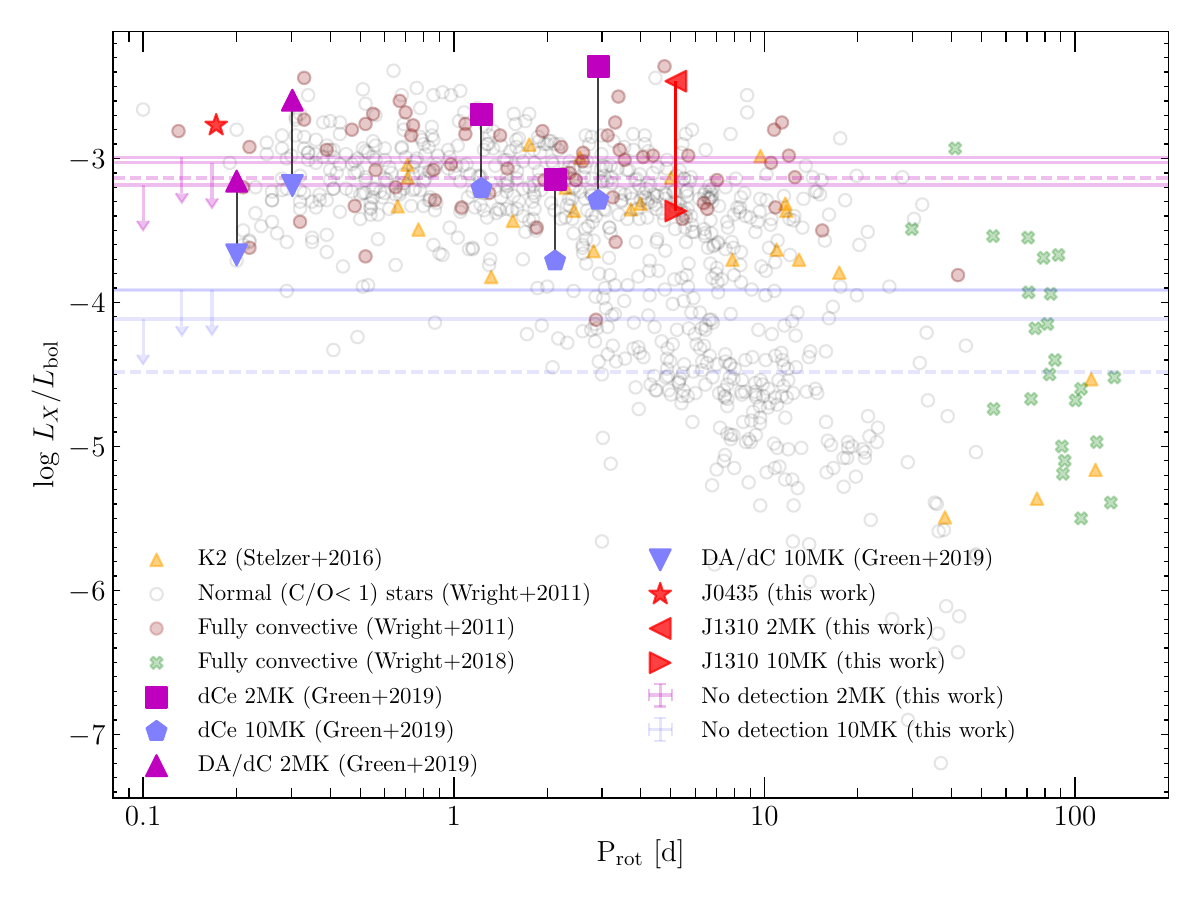}
\caption{Updated activity-rotation figure from \citet{Green2019}. For context, the activity and rotation for normal (C/O~$< 1$) main-sequence stars are shown in the background. The logarithm of X-ray to bolometric luminosity is plotted against rotation period for the normal dwarf stars from the samples of: \citet{Wright2011}, \citet{Stelzer2016}, and \citet{Wright2018}. We show the dCs from the \citet{Green2019} \Chandra\ sample and use the recently found rotation periods from \citet{Roulston2021a} and \citet{Whitehouse2021} to place those dCs at their true location on the diagram, assuming both 2MK and 10MK APEC plasma models for X-ray emission.  The X-ray detected dC J1548 from the \citet{Green2019} sample does not have a rotation period and so is marked by horizontal dashed lines. Finally, we show the two dCs with new X-ray detections reported in this paper, J0435 and SBSS1310, as well as the $3\sigma$ upper limits for the three non-detections. The rotation period for SBS1310 comes from \citet{Roulston2021a}, and the rotation period for J0435 is from this paper. For J0435, we use the simultaneous best-fit X-ray model of 13.4MK (section \ref{sec:J0435_indfit}).  It is clear that for either $2$\,MK and $10$\,MK models, most dCs are found in the saturated regime of active stars.}
\label{fig:Prot_vs_Lx}
\end{figure*}

With the newly found periods for the dCs in \citet{Green2019} and this work, we can now place dCs on an activity-rotation diagram. Figure \ref{fig:Prot_vs_Lx} shows the updated Figure 3 of \citet{Green2019}, but we now show the true position of the five dCs in that sample with their rotation periods (the one dC from that sample without a rotation period is still shown with horizontal lines). We additionally place the two new X-ray detected dCs from this work, SBSS1310 and J0435, on this diagram. For SBSS1310 we include both the 2\,MK and 10\,MK model values, both placing SBSS1310 in the saturated regime. For J0435, we use the simultaneous fit model, making it the best-constrained dC of both samples. J0435 is clearly in the saturated regime with a short rotation period ($P=0.1719$\,d) and $\log{(\LxLbol)} = -2.77$.

The dC sample of \citet{Green2019} was chosen because its stars were \textit{expected} to show X-ray activity due to being either composite spectroscopic binaries of the dC+DA type, or showing signs of chromospheric activity with H$\alpha$ emission. This suggested that dCs have experienced spin-up from the angular momentum of the accreted material during mass transfer. As dCs are believed to be from older thick disk and halo populations \citep{Green2013, Farihi2018}, we should expect them to have spun-down from magnetic braking and angular momentum loss through magnetized winds \citep{Kraft1967, Matt2015, Garraffo2018}; therefore, any signs of short rotation periods would be indicators of mass transfer spin-up. Indeed, the location of the dCs from \citet{Green2019} in the rotation-activity diagram indicated short rotation periods, which have now been confirmed by \citet{Roulston2021a} and \citet{Whitehouse2021}. 

The dCs in the current study were selected to investigate if dCs, regardless of H$\alpha$ emission or a spectroscopically detectable WD, show signs of spin-up and chromospheric rejuvenation. Indeed, we find that the two dCs with X-ray detections in this work are both in the saturated activity regime with short rotation periods. 

However, the recent works of \citet{Roulston2021a} and \citet{Whitehouse2021}, where  a large number of new dC periods were found, complicate the interpretation of activity as resulting only from accretion-induced spin-up \citep{Green2019}. Remarkably, 95\% of the new dC periods are under $10$\,d, with nine having been confirmed to have the same photometric (likely rotational) period and orbital period. Since dCs form via mass transfer from evolved TP-AGB stars and TP-AGB stars can reach radii of 800\,R$_\odot$ (3.7\,au) as they experience successively stronger thermal pulses \citep{Marigo2017}, these short period dC stars must have experienced a common-envelope phase \citep{Paczynski1976, Ivanova2013}. Therefore, the dCs in this paper, and in \citet{Green2019}, must have experienced a common-envelope \added{phase} and the associated spiral-in to these short periods. This spiral-in results in the circularization and subsequent synchronization of the binary \citep{Hurley2002}, and therefore the resulting final dC should have a rotation period commensurate with post-common-envelope binaries (PCEBs), i.e. P$\approx1$\,d. Thus, the X-ray detections in \citet{Green2019} and in this work do indeed trace short period rotation of dCs; however, the cause of this dC spin-up is more likely associated with common-envelope spiral-in, and subsequent spin-orbit locking in the binary system with the remnant WD, and not necessarily angular momentum gain from accreting carbon-rich material. A more appropriate \Chandra\ sample to probe accretion-induced spin-up would be to target dCs in which the orbital period is on the order of years.  For example, the 3 dCs from \citet{Harris2018} (with astrometric periods of  1.23\,yr, 3.21\,yr, and 11.35\,yr) should have avoided a common-envelope phase, and therefore, the rotation period should only have been affected by accretion. 

\section{J0435 Spectral Energy Distribution}\label{sec:J0435_SED}

The significant column density (N$_{\rm H} = 1.77\times10^{22}$\,cm$^{-2}$) from the spectral fit of J0435 indicates the presence of substantial material along our line of sight. However, the expected intervening column density from the Bayestar17 dust maps (N$_{\rm H} = 6.3\times10^{19}$\,cm$^{-2}$) suggest negligible amounts of dust in the direction and distance of J0435. This suggests that there may be either substantial circumbinary or circumstellar material around J0435. 

The mass transfer process to form dCs requires the accretion of carbon-rich material from a former TP-AGB companion (which now as a WD, has cooled beyond detection). The carbon-rich dust expelled by these TP-AGB stars has large opacity to optical and infrared photons, driving high radiation pressure and therefore large mass loss rates of $\sim 10^{-7}$--$10^{-5}$\,M$_\sun$\,yr$^{-1}$ \citep{Hofner2018}. This should result in extended shells of dust around nascent dC systems. In addition, those dCs that experience a common-envelope phase will eject the envelope of the TP-AGB star resulting in a planetary nebula. This seems to be observed in the Necklace Nebula where the central source was found to be a binary with a dC, having a photometric period of 1.16\,d \citep{Corradi2011, Miszalski2013}.
\added{As the WD companion to the newly minted dC cools, the planetary nebula should similarly fade (on typical timescales of $\sim 10^4$ years), but may leave detectable signs of circumbinary dust and gas around the dC. Given that dC main sequence lifetimes can exceed planetary nebula lifetimes by a factor of  $\sim 10^5$,  it is perhaps a surprise that even one dC is known within a cataloged PN.}

We compiled the spectral energy distribution (SED) of J0435 using a variety of catalog observations. In the optical, we cross-matched to \Gaia EDR3 \citep{GaiaEDR3} and the Pan-STARRS1 survey \citep{PanSTARRS1-1, PanSTARRS1-2, PanSTARRS1-3, PanSTARRS1-4, PanSTARRS1-5, PanSTARRS1-6}. In the near-infrared and mid-infrared, we cross-matched to the Two Micron All-Sky Survey \citep[2MASS;][]{2MASS} and \textit{WISE} surveys respectively. We also cross-matched to the GALEX GR6/7 \citep{GALEX} finding only a near-ultraviolet (NUV, 130–180 nm) detection for J0435.

We obtained deeper NUV and far-ultraviolet (FUV) observations of J0435 using the Wide Field Camera 3 (WFC) and Advanced Camera for Surveys (ACS) detectors on the \textit{Hubble Space Telescope} (HST). We obtained NUV images with WFC3 using the F225W filter, across one full orbit with a total exposure time of 2384.0s. The exposure was split into four equal exposures of 596.0s and dithered using the WFC3-UVIS-DITHER-BOX pattern, with point spacing of 0.173\arcsec\ and line spacing of 0.112\arcsec. We obtained FUV images with the Solar Blind Channel (SBC) on the ACS using both the F140LP and F165P filters\footnote{We obtained FUV images in both filters to account for the SBC red leak.}. Observations with both filters were obtained within one orbit, with a total exposure of 1084.0s and 1091.0s in the F140LP and F165LP filters respectively. Exposures in both filters were split into four equal exposures (271.0s, 272.75s) and dithered using the ACS-SBC-DITHER-BOX pattern, with point spacing of 0.179\arcsec\ and line spacing of 0.116\arcsec.

We measured the NUV and FUV magnitudes in a similar way. We used the \Gaia EDR3 positions and proper motions to update the coordinates of J0435 to the time of the \textit{HST} observations. We performed simple aperture photometry using a circular aperture (with radius 0.8\arcsec\ and 0.2\arcsec\ for the NUV/FUV images respectfully) for the source and an annulus (with inner and outer radii of 2\arcsec\ and 4\arcsec\ for the NUV and 0.5\arcsec\ and 4.5\arcsec\ for the FUV) for the background region. For both the NUV and FUV images, we correct the aperture counts using the provided encircled energy fractions\footnote{\url{https://www.stsci.edu/hst/instrumentation/wfc3/data-analysis/photometric-calibration/uvis-encircled-energy}}\footnote{\url{https://www.stsci.edu/hst/instrumentation/acs/data-analysis/aperture-corrections}}. The measured NUV (F225W) magnitude is $22.224\pm0.003$, and the measured FUV (1400\AA--1650\AA) magnitude is extremely faint, at $29.04\pm0.95$.

\begin{figure*}
\centering
\plotone{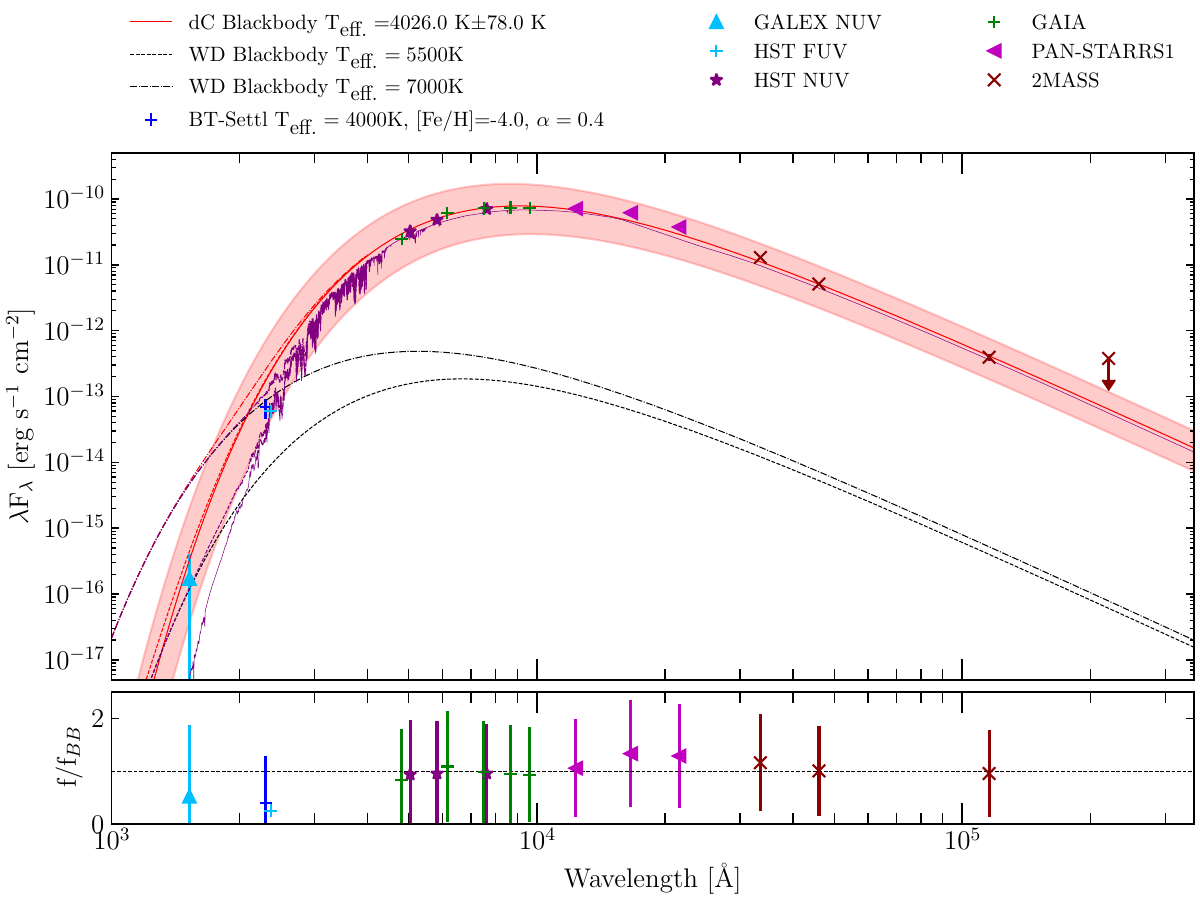}
\caption{Spectral energy distribution for the dC J0435. The SED was compiled from \Gaia EDR3 and Pan-STARRS1 in the optical, GALEX in the NUV, 2MASS in the near-infrared, \textit{WISE} in the mid-infrared, and new \textit{HST} observations in the NUV/FUV regions. The best fitting blackbody model is shown as the solid red line and consists of a $4026$\,K$\pm78$\,K dC with a radius of $0.48$\,R$_\sun\pm0.15$\,R$_\sun$. The red shaded region shows the $1\sigma$ uncertainty of this best-fit model. The bottom panel shows the observed flux divided by the expected flux from the best-fit blackbody for each observation; error bars represent the blackbody fit errors added in quadrature to the photometric flux uncertainties. Also shown is a BT-Settl model atmosphere \citep{BTSettl} with [Fe/H]$=-4.0$ and $\alpha=0.4$, normalized to the \Gaia EDR3 distance and fit radius (again with the shaded purple region showing the $1\sigma$ uncertainty). The black dashed and dotted-dashed lines show two blackbodies of a $7000$\,K and $5500$\,K respectively meant to describe the WD. For each WD model, the combined dC blackbody and combined BT-Settl models are shown. The new \textit{HST} fluxes are consistent with a $4000$\,K main-sequence star combined with a cool WD. The FUV flux places a weak constraint of approximately $5500$\,K on the WD companion. In addition, there appears to be a slight increase in the near-infrared flux as compared to the best-fit blackbody, but within the $1\sigma$ uncertainty region. }
\label{fig:J0435_SED}
\end{figure*}

Figure \ref{fig:J0435_SED} shows the SED of J0435 with the catalog and new \textit{HST} fluxes. We fit a blackbody model to the SED of J0435, excluding the \textit{HST} and GALEX FUV/NUV fluxes. We corrected for the expected extinction by using the Bayestar17 dust maps \citep{Bayestar2017}. We used the extinction law from \citet{Cardelli1989}, assuming $R_V = 3.1$, to calculate the extinction in the observed bands. The fit results in a dC temperature of $4026$\,K$\pm78$\,K and dC radius of $0.48$\,R$_\sun\pm0.15$\,R$_\sun$. This best-fit model is shown as the solid red line in Figure \ref{fig:J0435_SED}, with the shaded red region representing the $1\sigma$ uncertainty region from this fit. Also shown (as the solid purple line) is a $4000$\,K BT-Settl model atmosphere \citep{BTSettl} with [Fe/H]$=-4.0$ and $\alpha=0.4$, normalized to the \Gaia EDR3 distance and fit radius. Figure \ref{fig:J0435_SED} also shows two blackbodies for a WD of $7000$\,K and $5500$\,K. For both WD blackbodies, we show the combined dC blackbody and dC BT-Settl model atmosphere as dashed lines.   

The dC blackbody model fit reproduces the observed dC SED moderately well, with the BT-Settl model atmosphere matching more closely in the UV region. Our measured \textit{HST} NUV flux is consistent with that of the GALEX NUV flux, and the flux from a $4000$\,K main-sequence BT-Settl model. Although our \textit{HST} flux has a fairly large uncertainty, it is consistent with the fit dC temperature and a cool WD companion. From Figure \ref{fig:J0435_SED}, it is clear the WD must be cooler than $7000$\,K or we would have observed a slightly higher NUV flux, and our FUV detection should have been much stronger. If we use our FUV flux as an upper limit of the WD flux contribution, we find the WD is likely around $5500$\,K implying a cooling age (therefore the time since mass transfer to the dC) of approximately $3.5$\,Gyr (using a standard WD mass of $0.6$\,M$_\odot$), assuming that there have been no accretion episodes since.

Circumbinary and circumstellar dust around J0435 should re-emit absorbed radiation in the mid-infrared. This should be visible as a bump in the infrared region of the SED. While there does appear to be a slight bump in the SED of J0435 in the 2MASS fluxes, it is within the $1\sigma$ uncertainties of the blackbody fit, supporting the non-detection of a dusty disk in the SED of J0435. Additionally, the SED is well fit using the negligible extinction from the Bayestar17 dustmaps, pointing again to a lack of dust along the line of sight to J0435.

\section{Discussion}\label{sec:discussion}

\citet{Green2019} sought to determine if dCs, while expected to be of older thick disk and halo populations \citep{Green2013, Farihi2018}, may still show signs of coronal activity due to rapid rotation induced by an increase in angular momentum from mass transfer. While they did indeed find that all of their observed dCs were detected with \Chandra\ and consistent with saturated X-ray activity, their pilot sample was biased to enhance detection probability, targeting dCs showing $H\alpha$ emission, a known tracer of coronal activity. 

Following up on successful detection of those dCs, our sample in this paper aimed to study a more representative sample, targeting the five nearest known dCs regardless of H$\alpha$ emission. Of the five dCs targeted, we detect X-ray emission in two. We use the same assumed 2\,MK and 10\,MK plasma temperature models to calculate $\log{(\LxLbol)}$, finding that both of those dCs fall in the saturated regime. For the three non-detections, we place $3\sigma$ upper level constraints on the X-ray flux.

For the dC J0435, we have seven individual \Chandra\ observations, with a total of 289 counts. This allowed us to fit the X-ray spectrum, placing constraints on the plasma temperature (T$_{\rm X} = 14.2$\,MK) and column density (N$_{\rm H} = 1.77\times10^{22}$\,cm$^{-2}$). The column density suggests a large amount of material surrounding J0435, but the lack of a mid-infrared excess in the SED, and the good SED fit without the need for a larger extinction correction, suggests that the material around J0435 may be gas with very little dust. The material may indeed be the remnants of the TP-AGB wind or common-envelope ejecta. This explanation is problematic, however, as the CE material would be expected to should have been cleared from the system, especially given the time since the CE inferred from the estimated cooling age. There could be the remains of the AGB wind, the CE, or an accretion disk that was in the form of dust but has been heated above the sublimation temperature by the strong X-ray activity found in J0435. The origin of this anomalous column density and low redenning motivates further multi-wavelength studies of J0435. 

While our results here are consistent with \citet{Green2019}, recent works have shown that the previous interpretation of dC X-ray activity, as primarily due to accretion spin-up,  may not be complete. \citet{Roulston2021a} and \cite{Whitehouse2021} recently found 40 dCs with photometric periods, with 95\% having P$<10$\,d. These dCs must have been engulfed in a common-envelope during their former giant companion's TP-AGB phase. This would have caused a spiral-in of the dC, after which tidal spin-orbit synchronization would lead to the observed short dC rotation periods. It appears that, compared to accretion-induced spin up as originally postulated by  \citet{Green2019}, these spin-orbit-induced short rotation periods are more likely the source of the increased coronal activity as traced by the X-ray emission. 

An interesting comparison to make is to the symbiotic stars \citep{Davidsen1976, Allen1984, Luna2013}. Symbiotic stars consist of a compact object in a bound orbit around a red giant and accreting from its wind. They are known to have orbital periods ranging from hundreds of days to thousands of days \citep{Mikolajewska2012}. The accretion in symbiotics is believed to take place via wind accretion or a form of WRLOF \citep{Luna2018}, both of which likely form an accretion disk around the compact object. In symbiotics with a WD, which are analogous to dCs, this accretion disk results in X-ray emission, with thermal bremsstrahlung models of $\sim 100$\,MK \citep{Chernyakova2005, Tueller2005, Mukai2007, Smith2008, Kennea2009, Luna2013, Luna2018, Danehkar2021}, compared to the approximate $\sim10$\,MK we find for the dCs with X-ray detections. This supports our conclusion that the observed X-ray emission in dCs is indeed from coronal activity and not from accretion onto the WD companion.

The question remains though of how the initial properties of both the binary and the individual stars affect the formation of dCs. The evolution of the TP-AGB star, and the subsequent third dredge-up events, are affected by both the initial mass and metallicity of the star \citep{Kalirai2014}; this includes the final C/O of the TP-AGB envelope, controlling the C budget available to enhance the proto-dC. If the initial orbital period is too short, the system risks entering a CE phase either during the red giant branch or during the AGB phase before the third dredge-up can enhance the AGB to C/O$ > 1$. If the initial orbital period is too long, than mass transfer may only take place via BHL accretion, or WRLOF may not effectively shrink the orbit to begin a CE, which would then cause the binary to spiral in to the observed short periods. Therefore, the initial orbital and stellar properties that can result in a dC, and more strictly short period dCs, must inhabit a parameter space more stringent than traditional (C/O $< 1$) WD+MS PCEBs, although they remain unknown.

\citet{Roulston2021a} examined whether main-sequence companions to TP-AGB stars can accrete enough mass during the common-envelope phase to form dCs. They found that the common-envelope efficiency must be low to account for the known short period orbits of dCs, which is consistent with the more well known normal (C/O $< 1$) WD+MS PCEBs \citep{Zorotovic2010, Toonen2013, Camacho2014}. Furthermore, they also found that dCs cannot accrete enough carbon rich material during the common-envelope phase, at least on the approximately 100\,yr common-envelope timescale assumed. They suggest that the dCs must accrete enough carbon rich material before the common-envelope \added{phase}, but after the third dredge-up has polluted the AGB companion, via WRLOF\citep{Mohamed2007}. In WRLOF, the primary (in the case of dCs, this would be a TP-AGB star) does not completely fill its Roche-Lobe, and the primary wind is focused in the orbital plane towards the secondary star (the proto-dC). This results in accretion rates which can be significantly higher than those in the Bondi-Hoyle-Lyttleton prescription, in some cases as high as 50\% \citep{Abate2013, Saladino2018, Saladino2019a, Saladino2019b}. It has also been shown that WRLOF can efficiently tighten the orbit \citep{Saladino2018, Chen2018}, driving these systems toward the short periods that have been found for dCs.

The WRLOF formalism for forming dCs requires a balance of initial orbital period, progenitor TP-AGB mass and metallicity, and progenitor dC mass and metallicity (and likely other parameters as well). 
It has been suggested that dCs may be predominantly found in low metallicity populations, as the amount of carbon excess needed to be accreted to make C/O$>1$ is less in a low metallicity star.  The prototype dC G77-61 is extremely metal deficient with [Fe/H]$\sim -4$ \citep{Plez2005}. The mass of the dC progenitor (and C/O of the accreted mass) will also change how much material must be accreted. \citet{Miszalski2013} estimated that to shift a secondary from (C/O)$_i \sim 1/3$ to (C/O)$_f > 1$ would require the accretion of $\Delta M_2 = 0.03-0.35$\,M$_\sun$ for a secondary with a mass $M_2 = 1.0-0.4$\,M$_\sun$. 
The TP-AGB phase can last from 1\,Myr up to 3.5\,Myr \citep{Kalirai2014}, while the C-AGB phase itself only lasts up to $\sim 0.42$\,Myr for an initial mass of 2.60\,M$_\sun$. Mass transfer to the dC must happen during this short time, which supports the WRLOF scenario, as a dC may accrete 0.35\,M$_\sun$ in only $10^3$--$10^6$\,yrs (for the above AGB mass loss rates of $10^{-7}$--$10^{-5}$\,M$_\sun$\,yr$^{-1}$,  \citealt{Hofner2018}) via WRLOF with accretion efficiencies as high as $\sim 50$\% \citep{Abate2013}. 


Systems in which the dC has not experienced a common-envelope \added{phase} (such as the dCs with orbital periods of a year or more \citep{Harris2018}, may be the best candidates yet for testing if the accretion of carbon-rich material can cause the rejuvenation of dCs via spin-up to short rotation periods. Additionally, future simulations of WRLOF and common-envelope evolution in progenitor dC systems, coupled with the observed dC space density and fraction of dCs with short orbital periods, may allow the first insight into the initial conditions needed for dC formation. 


\facility{CXO (ACIS-S), HST (WFC3, ACS), 2MASS, \textit{Gaia}, \textit{WISE},  \textit{TESS}}

\software{Astropy \citep{astropy1, astropy2}, CIAO \citep{CIAO}, dustmaps \citep{dustmaps}, Matplotlib \citep{matplotlib}, Numpy \citep{numpy}, Scipy \citep{scipy}, Sherpa \citep{Sherpa}, TOPCAT \citep{topcat}}

\begin{acknowledgments}
B.R. was supported for this work  by the National Aeronautics and Space Administration through Chandra Award Numbers GO0-21003X and GO1-22004X issued by the Chandra X-ray Center, which is operated by the Smithsonian Astrophysical Observatory for and on behalf of the National Aeronautics Space Administration under contract NAS8-03060.  Additional support for observations made with the NASA/ESA Hubble Space Telescope was provided under program number GO-16392, provided by NASA through a grant from STScI, which is operated by AURA, Inc., under NASA contract NAS 5-26555. 
\end{acknowledgments}
\begin{acknowledgments}
We acknowledge the use of TESS High Level Science Products (HLSP) produced by the Quick-Look Pipeline (QLP) at the TESS Science Office at MIT, which are publicly available from the Mikulski Archive for Space Telescopes (MAST). Funding for the TESS mission is provided by NASA's Science Mission directorate.
\end{acknowledgments}
\begin{acknowledgments}
Funding for the Sloan Digital Sky Survey IV has been provided by the Alfred P. Sloan Foundation, the U.S. Department of Energy Office of Science, and the Participating Institutions. 

SDSS-IV acknowledges support and resources from the Center for High Performance Computing  at the University of Utah. The SDSS website is www.sdss.org.

SDSS-IV is managed by the Astrophysical Research Consortium for the Participating Institutions of the SDSS Collaboration including the Brazilian Participation Group, the Carnegie Institution for Science, Carnegie Mellon University, Center for Astrophysics $|$ Harvard \& Smithsonian, the Chilean Participation Group, the French Participation Group, Instituto de Astrof\'isica de Canarias, The Johns Hopkins University, Kavli Institute for the Physics and Mathematics of the Universe (IPMU) / University of Tokyo, the Korean Participation Group, Lawrence Berkeley National Laboratory, Leibniz Institut f\"ur Astrophysik Potsdam (AIP),  Max-Planck-Institut f\"ur Astronomie (MPIA Heidelberg), Max-Planck-Institut f\"ur Astrophysik (MPA Garching), Max-Planck-Institut f\"ur Extraterrestrische Physik (MPE), National Astronomical Observatories of China, New Mexico State University, New York University, University of Notre Dame, Observat\'ario Nacional / MCTI, The Ohio State University, Pennsylvania State University, Shanghai Astronomical Observatory, United Kingdom Participation Group, Universidad Nacional Aut\'onoma de M\'exico, University of Arizona, University of Colorado Boulder, University of Oxford, University of Portsmouth, University of Utah, University of Virginia, University of Washington, University of Wisconsin, Vanderbilt University, and Yale University.
\end{acknowledgments}

\bibliography{references}{}
\bibliographystyle{aasjournal}



\end{document}